\documentclass[aps,pra,twoside,twocolumn,a4paper,reprint,noeprint]{revtex4-2}
\usepackage[utf8]{inputenc}
\usepackage{graphicx}
\usepackage{amsmath}
\usepackage{amsfonts}
\usepackage{amssymb}
\usepackage{mathtools}
\usepackage[usenames,dvipsnames,svgnames]{xcolor}
\usepackage[USenglish]{babel}
\usepackage{hyperref}
\usepackage{grffile}
\usepackage{cleveref}
\Crefname{figure}{Fig.}{Figs.}
\usepackage{multirow}

\usepackage{array}

\allowdisplaybreaks
\begin{document}

\title{Complete zero-energy flat bands of surface states in\\ fully gapped chiral noncentrosymmetric superconductors}

\author{Clara J. Lapp}
\email{clara{\_}johanna.lapp@tu-dresden.de}
\affiliation{Institute of Theoretical Physics, Technische Universit\"at Dresden, 01069 Dresden, Germany}
\affiliation{W\"urzburg--Dresden Cluster of Excellence ct.qmat, Technische Universit\"at Dresden, 01062 Dresden, Germany}

\author{Julia M. Link}
\email{julia.link@tu-dresden.de}
\affiliation{Institute of Theoretical Physics, Technische Universit\"at Dresden, 01069 Dresden, Germany}
\affiliation{W\"urzburg--Dresden Cluster of Excellence ct.qmat, Technische Universit\"at Dresden, 01062 Dresden, Germany}

\author{Carsten Timm}
\email{carsten.timm@tu-dresden.de}
\affiliation{Institute of Theoretical Physics, Technische Universit\"at Dresden, 01069 Dresden, Germany}
\affiliation{W\"urzburg--Dresden Cluster of Excellence ct.qmat, Technische Universit\"at Dresden, 01062 Dresden, Germany}

\begin{abstract}
Noncentrosymmetric superconductors can support flat bands of zero-energy surface states in part of their surface Brillouin zone. This requires that they obey time-reversal symmetry and have a sufficiently strong triplet-to-singlet-pairing ratio to exhibit nodal lines in the bulk. These bands are protected by a winding number that relies on chiral symmetry, which is realized as the product of time-reversal and particle-hole symmetry. We reveal a way to stabilize a flat band in the entire surface Brillouin zone, while the bulk dispersion is fully gapped. This idea could lead to a robust platform for quantum computation and represents an alternative route to strongly correlated flat bands in two dimensions, besides twisted bilayer graphene. The necessary ingredient is an additional spin-rotation symmetry that forces the direction of the spin-orbit-coupling vector not to depend on the momentum component normal to the surface.
We define a winding number that leads to flat zero-energy surface bands due to bulk-boundary correspondence. We discuss under which conditions this winding number is nonzero in the entire surface Brillouin zone and verify the occurrence of zero-energy surface states by exact numerical diagonalization of the Bogoliubov--de Gennes Hamiltonian for a slab. In addition, we consider how a weak breaking of the additional symmetry affects the surface band, employing first-order perturbation theory and a quasiclassical approximation.
We find that the surface states and the bulk gap persist for weak breaking of the additional symmetry but that the band does not remain perfectly flat. The broadening of the band strongly depends on the deviation of the spin-orbit-coupling vector from its unperturbed direction as well as on the spin-orbit-coupling strength and the triplet-pairing amplitude.
\end{abstract}

\maketitle

\section{Introduction}
\label{sec:intro}

Noncentrosymmetric superconductors with time-re\-ver\-sal symmetry (TRS) have recently attracted a lot of attention. For sufficiently strong spin-triplet contribution to the superconducting pairing, these materials posses nodal lines in their bulk dispersion, which are associated with a winding number. At their surface, they exhibit zero-energy Majorana states in the part of the surface Brillouin zone (sBZ) that is enclosed by the projection of the nodal lines. The resulting zero-energy flat bands are topologically protected by the winding number of the bulk nodal lines \cite{SR11, BST11, STY11, SBT12, HQS13, SB15}. These flat bands of Majorana modes are of particular interest in the context of quantum computation \cite{K03, NSSFS08, SLTD10, LSD10, ORO10, BTS13, EF15, SFN15, OO20, RRT20, LT22}.

Majorana modes are attractive for quantum computation on the one hand because qubit states realized by Majorana modes are topologically protected, which promises high stability against relaxation and decoherence. On the other hand, quantum gates can be realized by moving such Majorana modes around each other, i.e., by braiding their world lines. There are excellent reviews covering these ideas  \cite{NSSFS08, SFN15}. Flat Majorana surface bands are intriguing in this context because they realize a \emph{macroscopic} number of localized Majorana modes. Observation of non-trivial braiding statistics would be a smoking-gun experiment for Majorana modes but the required manipulation of individual Majorana modes is likely difficult~\cite{LT22}.

Another interesting aspect is that the flat bands are automatically in the strong-coupling regime since their kinetic energy is zero \cite{PoL14, CPF15, RRT20}. A similar situation with flat bands of Majorana modes within the projection of nodal lines has also been discussed in the context of inversion-symmetric superconductors, where the nonzero winding number arises from an orbital degree of freedom that transforms nontrivially under time reversal \cite{WW23}. However, near the boundary of the flat-band region of the nodal systems, the bulk-state energies come arbitrarily close to the surface band due to the gap closing in the bulk, which could hinder the experimental detection of the flat band and compromise the robustness of quantum-computation schemes. Importantly, the winding number in these systems has to vanish in a finite fraction of the sBZ. For example, due to TRS, the region with a nonzero winding number never includes the origin. The reason is that every nodal line has a time-reversed partner. Therefore, if the projection of one of the nodal lines encloses the origin, the projection of its time-reversed partner does so as well. The two lines lead to opposite winding numbers, which compensate each other in the intersecting region so that the winding number at the origin is zero. To avoid this problem, one would have to break TRS. However, the winding number and therefore the topological protection of the surface states relies on the presence of a chiral symmetry, which, in these systems, is realized as the product of TRS and particle-hole symmetry (PHS). Since the latter is always present in the Bogoliubov--de Gennes (BdG) formalism, breaking TRS destroys the topological protection.

It would be highly desirable to energetically separate the flat surface band from the bulk states by having a full bulk gap and a topologically protected zero-energy flat surface band in the entire sBZ. Even in the presence of terms that weakly break the symmetries which protect the zero-energy states, this could still lead to an approximately flat surface band close to zero energy within a full bulk gap. One possible route would be a system without TRS but with a chiral symmetry arising from a sublattice symmetry, i.e., the system can be decomposed into two sublattices and there only is hopping between these sublattices and not within them. However, we do not concentrate on this idea since beyond-nearest-neighbor hopping terms destroy the chiral symmetry. Another situation where the occurrence of flat bands in the full sBZ due to chiral symmetry has been analyzed are crystalline topological insulators \cite{NHV21}. Here, the protection of the boundary modes arises from geometric deformations and leads to a finite polarization of the surface state.

In this paper, we instead present an approach to realize a similar type of winding number as in the nodal system---but in the entire sBZ---by introducing an additional symmetry. Specifically, we require the existence of a spin component that commutes with the Hamiltonian. This requires the direction of the spin-orbit-coupling (SOC) vector not to depend on the momentum component normal to the surface and allows us to bring the Hamiltonian into block-diagonal form. The most straightforward possibility is for the SOC vector to point in a single direction for all momenta. We find that for all point groups except for the three low-symmetry groups $C_1$, $C_2$, and $C_s$, this is the only possibility, if we restrict ourselves to the lowest-order expansion of the SOC vector. The resulting spin-up and spin-down blocks no longer obey TRS and PHS separately, which are both present in the BdG Hamiltonian of the full system. However, the chiral symmetry persists in the spin blocks, i.e., there are unitary matrices that anticommute with the spin-up and spin-down blocks. Hence, the individual blocks are in symmetry class AIII \cite{SRFL08,AZ97,Z96,SBT12}. Relying on the chiral symmetry, we can derive a momentum-dependent winding number in the sBZ for both the spin-up and the spin-down block. Each of these winding numbers protects a flat band of zero-energy surface states, leading to a twofold degenerate flat surface band in the entire sBZ, while the bulk is fully gapped.

We discuss the conditions under which these winding numbers are nonzero and verify our results numerically through exact diagonalization of the BdG Hamiltonian of a slab. Moreover, we consider the effects of a weak breaking of the spin-component symmetry, i.e., an only approximately unidirectional SOC vector. We use a quasiclassical approximation for the surface states in the direction orthogonal to the surface and derive the first-order correction to the flat-band energy within perturbation theory. We compare the results to exact diagonalization of the BdG Hamiltonian.

The remainder of this paper is organized as follows: In Sec.~\ref{sec:Model system, symmetries and winding number}, we introduce a model Hamiltonian, discuss its symmetries, and derive the winding number. We discuss the necessary conditions for a nonzero winding number that protects surface states at zero energy. In Sec.~\ref{sec:ideal_system}, we list the crystallographic point groups which can in principle satisfy all necessary symmetries. Moreover, we determine the parameter regime that allows for a nonzero winding number in the entire sBZ. We demonstrate the occurrence of zero-energy flat bands for one exemplary system. In Sec.~\ref{sec:nonideal_systems}, we consider the implications of an imperfectly unidirectional SOC vector. Finally, in Sec.~\ref{sec:summary}, we summarize our findings and draw conclusions.

\section{Model system} \label{sec:Model system, symmetries and winding number}

We consider a three-dimensional noncentrosymmetric time-reversal-symmetric single-band superconductor modeled by a Hamiltonian
\begin{equation}\label{eq:hamiltonian}
\mathcal{H}=\frac{1}{2}\sum_{\mathbf{k}} \Psi_{\mathbf{k}}^\dagger \mathcal{H}(\mathbf{k})\Psi_{\mathbf{k}},
\end{equation}
with $\Psi_{\mathbf{k}}=(c_{\mathbf{k},\uparrow},c_{\mathbf{k},\downarrow}, c_{-\mathbf{k},\uparrow}^\dagger,c_{-\mathbf{k},\downarrow}^\dagger)^T$, where $c^\dagger_{\mathbf{k}\sigma}$ ($c_{\mathbf{k}\sigma}$) is the creation (annihilation) operator of an electron with wave vector $\mathbf{k}$ and spin $\sigma\in\lbrace\uparrow,\downarrow\rbrace$, and the BdG Hamiltonian
\begin{equation}\label{eq:bdg_hamiltonian}
\mathcal{H}(\mathbf{k})=\begin{pmatrix}
h(\mathbf{k})& \Delta(\mathbf{k})\\
\Delta^\dagger(\mathbf{k})&-h^T(-\mathbf{k})
\end{pmatrix}.
\end{equation}
The matrix $h(\mathbf{k})$ is the normal-state Hamiltonian
\begin{equation}
h(\mathbf{k})=\epsilon_{\mathbf{k}} \sigma_0 + \mathbf{g}_{\mathbf{k}} \cdot \boldsymbol{\sigma},
\end{equation}
where the spin-independent part $\epsilon_{\mathbf{k}}$ is an even function of momentum $\mathbf{k}$, while the SOC vector $\mathbf{g}_{\mathbf{k}}$ is odd in $\mathbf{k}$, and $\boldsymbol{\sigma}$ and $\sigma_0$ represent the vector of Pauli matrices and the $2 \times2$ identity matrix, respectively.
Due to the breaking of inversion symmetry in the normal state, the superconducting pairing matrix $\Delta(\mathbf{k})$ generically contains both a singlet component, which is even in $\mathbf{k}$, and a triplet component, which is odd in $\mathbf{k}$, giving
\begin{equation}\label{eq:gap_matrix}
\Delta(\mathbf{k})=(\Delta^s_\mathbf{k} \sigma_0 + \mathbf{d}_\mathbf{k} \cdot \boldsymbol{\sigma})\,
  i \sigma_y ,
\end{equation}
where $\Delta^s_\mathbf{k}$ is the singlet pairing amplitude and $\mathbf{d}_\mathbf{k}$ is the triplet pairing vector. In the following, we will assume $\mathbf{d}_\mathbf{k} \parallel \mathbf{g}_{\mathbf{k}}$ as this state is known to be the most stable in the absence of interband pairing \cite{FAKS04,BST11}. We thus write $\mathbf{d}_{\mathbf{k}}=\Delta^t_\mathbf{k} \mathbf{l}_{\mathbf{k}}$ and $\mathbf{g}_{\mathbf{k}}=\lambda \mathbf{l}_{\mathbf{k}}$ with the triplet pairing amplitude $\Delta^t_\mathbf{k}$ and the SOC strength $\lambda$.
In this paper, we consider pairing that does not break the lattice symmetry, i.e., that transforms according to the trivial irreducible representation of the point group. This requires $\Delta^s_\mathbf{k}$ and $\Delta^t_\mathbf{k}$ to have the full symmetry of the normal-state Hamiltonian.

We are interested in superconductors with a full bulk gap. The simplest fully gapped state is realized for $(s+p)$-wave pairing with $\Delta^s = \mathrm{const}$ and $\Delta^t = \mathrm{const}$, which we will assume for our numerical calculations. However, our theory applies whenever there is a full bulk gap. Conversely, if the bulk gap has nodes, the winding number that we will introduce in Sec.~\ref{subsec:winding} is ill defined at the projections of these nodes into the sBZ.

\subsection{Symmetries}
\label{sub:symmetries}

A fully gapped Hamiltonian of the form described above belongs to the symmetry class DIII of the tenfold-way classification \cite{SRFL08,AZ97,Z96,SBT12,SB15} as it obeys both TRS $\mathcal{T}$ with $\mathcal{T}^2=-1$ and PHS $\mathcal{C}$ with $\mathcal{C}^2=+1$.
Specifically, the BdG formalism enforces $\mathcal{C}=\mathcal{K} U_C$ with $U_C=\sigma_x \otimes \sigma_0$ such that
\begin{equation}
U_C \mathcal{H}(-\mathbf{k})^T U_C^\dagger=-\mathcal{H}(\mathbf{k}).
\end{equation}
TRS can be written as $\mathcal{T}=\mathcal{K} U_T$ with $U_T=\sigma_0 \otimes i \sigma_y $ such that
\begin{equation}
U_T \mathcal{H}(-\mathbf{k})^T U_T^\dagger= +\mathcal{H}(\mathbf{k}).
\end{equation}
The combination of these two antiunitary symmetries gives the so-called chiral symmetry
\begin{equation}
U_S^\dagger \mathcal{H}(\mathbf{k}) U_S= -\mathcal{H}(\mathbf{k}),
\end{equation}
i.e., there is a unitary matrix $U_S=i U_T U_C=-\sigma_x \otimes \sigma_y$ that anticommutes with the Hamiltonian. Note that half of the eigenvalues of $U_S$ are equal to $+1$ and the other half to $-1$.

In addition to TRS, PHS, and chiral symmetry, we require the crystal structure to belong to a noncentrosymmetric point group. Every element of the point group represented by a $3\times 3$ orthogonal matrix $R$ leads to a relation~\cite{SBT12}
\begin{equation}
U_{\tilde{R}} \mathcal{H}(R^{-1}\mathbf{k}) U_{\tilde{R}}^\dagger= \mathcal{H}(\mathbf{k})
\end{equation}
satisfied by the Hamiltonian, where $\tilde{R} = R/\det(R) = \det(R) R$, i.e., $\tilde{R}$ describes a proper rotation by an angle $\theta$ about an axis denoted by a unit vector $\mathbf{n}$, and $U_{\tilde{R}}=\text{diag}( u_{\tilde{R}},u_{\tilde{R}}^\ast)$ with the spinor representation $u_{\tilde{R}}=\exp [-i\theta(\mathbf{n} \cdot \boldsymbol{\sigma})/2]$ of $\tilde{R}$.
This leads to the restrictions
\begin{align}
\mathbf{l}_\mathbf{k} &= \tilde{R}\, \mathbf{l}_{R^{-1}\mathbf{k}}, \label{eq:group_relation_l}\\
\epsilon_\mathbf{k} &= \epsilon_{R^{-1}\mathbf{k}},\\
\Delta^s_\mathbf{k} &= \Delta^s_{R^{-1}\mathbf{k}}.
\end{align}

We will need to fix a surface orientation, which gives one direction $k_\perp$ in the Brillouin zone that is orthogonal to the surface and a vector $\mathbf{k}_\shortparallel$ that parameterizes the two directions of the sBZ so that each point $\mathbf{k}$ in the three-dimensional Brillouin zone can be identified by some pair $(k_\perp, \mathbf{k}_\shortparallel)$.

We now introduce an additional symmetry for the Hamiltonian $\mathcal{H}(\mathbf{k})$. This symmetry is the main ingredient that will enable us to define a winding number and obtain protected zero-energy surface bands. As the new symmetry, we require that for each point $\mathbf{k}_\shortparallel$ in the sBZ, there is a spin component $\Sigma_{\mathbf{n}_{\mathbf{k}_\shortparallel}}$ in some direction $\mathbf{n}_{\mathbf{k}_\shortparallel}$ that commutes with the Hamiltonian for all momenta $k_\perp$ in the orthogonal direction, i.e.,
\begin{equation}\label{eq:spin_symmetry}
[\mathcal{H}(k_\perp,\mathbf{k}_\shortparallel),
  \Sigma_{\mathbf{n}_{\mathbf{k}_\shortparallel}}] = 0 \quad \forall k_\perp\in [ -\pi,\pi),
\end{equation}
where $\Sigma_{\mathbf{n}_{\mathbf{k}_\shortparallel}}$ is defined as
\begin{equation}\label{eq:Sigma0}
\Sigma_{\mathbf{n}_{\mathbf{k}_\shortparallel}}=\begin{pmatrix}
\mathbf{n}_{\mathbf{k}_\shortparallel} \cdot \boldsymbol{\sigma} & 0\\ 0& -\mathbf{n}_{-\mathbf{k}_\shortparallel} \cdot \boldsymbol{\sigma}^*
\end{pmatrix} ,
\end{equation}
with a unit vector $\mathbf{n}_{\mathbf{k}_\shortparallel}=(n^x_{\mathbf{k}_\shortparallel}, n^y_{\mathbf{k}_\shortparallel}, n^z_{\mathbf{k}_\shortparallel})^T$.
While the condition in Eq.~\eqref{eq:spin_symmetry} does not impose any additional constraints on $\epsilon_\mathbf{k}$ and $\Delta^s_{\mathbf{k}}$, it does require the SOC vector to be parallel to the vector $\mathbf{n}_{\mathbf{k}_\shortparallel}$, i.e.,
\begin{equation}
\mathbf{l_k} = l(\mathbf{k})\, \mathbf{n}_{\mathbf{k}_\shortparallel}
  = l(k_\perp,\mathbf{k}_\shortparallel)\, \mathbf{n}_{\mathbf{k}_\shortparallel}
  \quad \forall k_\perp\in [ -\pi,\pi),
\end{equation}
with a real-valued function $l(\mathbf{k})$. While $l(\mathbf{k})$ is odd in $\mathbf{k}$, $\mathbf{n}_{\mathbf{k}_\shortparallel}$ does not depend on the component $k_\perp$ and is thus even in $k_\perp$. Hence, $l(k_\perp,\mathbf{k}_\shortparallel)$ has to be odd in $k_\perp$.
Note that $l(\mathbf{k})$ is not the norm of $\mathbf{l}(\mathbf{k})$. Since $\mathbf{n}_{\mathbf{k}_\shortparallel}$ must now be even, Eq.\ \eqref{eq:Sigma0} can be written as
\begin{equation}\label{eq:Sigma}
\Sigma_{\mathbf{n}_{\mathbf{k}_\shortparallel}}=\begin{pmatrix}
\mathbf{n}_{\mathbf{k}_\shortparallel} \cdot \boldsymbol{\sigma} & 0\\ 0& -\mathbf{n}_{\mathbf{k}_\shortparallel} \cdot \boldsymbol{\sigma}^*
\end{pmatrix} .
\end{equation}
To ensure uniqueness, we choose the sign of $l(\mathbf{k})$ such that $n^z_{\mathbf{k}_\shortparallel}\geq 0$.

While the occurrence of Eq.~\eqref{eq:spin_symmetry} as an exact physical symmetry is unlikely as it is not imposed by any point group, the equation can still be satisfied approximately. We will discuss in Sec.\ \ref{sec:ideal_system} for which point groups this can be expected to occur. In the present section, we assume the exact validity of Eq.~\eqref{eq:spin_symmetry}. The effects of small deviations from this symmetry are treated in Sec.~\ref{sec:nonideal_systems}.

As the matrix $\Sigma_{\mathbf{n}_{\mathbf{k}_\shortparallel}}$ has the two eigenvalues $\pm 1$ that are both twofold degenerate the Hamiltonian $\mathcal{H}(\mathbf{k})$ can be brought into block-diagonal form via the transformation
\begin{equation}
\tilde{\mathcal{H}}(\mathbf{k})=\begin{pmatrix}
\mathcal{H}_\uparrow(\mathbf{k}) & 0\\ 0 & \mathcal{H}_\downarrow(\mathbf{k})
\end{pmatrix}=W_\Sigma(\mathbf{k}_\shortparallel) \mathcal{H}(\mathbf{k}) W_\Sigma^\dagger(\mathbf{k}_\shortparallel) ,
\end{equation}
where $W_\Sigma(\mathbf{k}_\shortparallel)$ is the matrix that diagonalizes $\Sigma_{\mathbf{n}_{\mathbf{k}_\shortparallel}}$ as
\begin{equation}
W_\Sigma(\mathbf{k}_\shortparallel) \Sigma_{\mathbf{n}_{\mathbf{k}_\shortparallel}} W_\Sigma^\dagger(\mathbf{k}_\shortparallel) = \sigma_z \otimes \sigma_0.
\end{equation}
Note that $\sigma\in\{\uparrow,\downarrow\}$ in $\mathcal{H}_\sigma$ refers to the spin orientation relative to the quantization axis $\mathbf{n}_{\mathbf{k}_\shortparallel}$. Performing the diagonalization and using $||\mathbf{n}_{\mathbf{k}_\shortparallel}||^2=1$, we obtain
\begin{align}\label{eq:W_Sigma}
W_\Sigma(\mathbf{k}_\shortparallel)=\begin{pmatrix}
 \frac{n^x_{\mathbf{k}_\shortparallel}+i n^y_{\mathbf{k}_\shortparallel}}{ \sqrt{2-2 n^z_{\mathbf{k}_\shortparallel}}} & \sqrt{\frac{1-n^z_{\mathbf{k}_\shortparallel}}{2}} & 0 & 0 \\
 0 & 0 & -\frac{n^x_{\mathbf{k}_\shortparallel}-i n^y_{\mathbf{k}_\shortparallel}}{\sqrt{2 n^z_{\mathbf{k}_\shortparallel}+2}} & \sqrt{\frac{n^z_{\mathbf{k}_\shortparallel}+1}{2}} \\
 -\frac{n^x_{\mathbf{k}_\shortparallel}+i n^y_{\mathbf{k}_\shortparallel}}{ \sqrt{2 n^z_{\mathbf{k}_\shortparallel}+2}} & \sqrt{\frac{n^z_{\mathbf{k}_\shortparallel}+1}{2}} & 0 & 0\\
 0 & 0 & \frac{n^x_{\mathbf{k}_\shortparallel}-i n^y_{\mathbf{k}_\shortparallel}}{ \sqrt{2-2 n^z_{\mathbf{k}_\shortparallel}}} & \sqrt{\frac{1-n^z_{\mathbf{k}_\shortparallel}}{2}}
\end{pmatrix}
\end{align}
for $n^z_{\mathbf{k}_\shortparallel} \neq 1$, i.e., for a SOC vector which is not parallel to the $z$-axis. For the special case of $\mathbf{n}_{\mathbf{k}_\shortparallel} =\hat{\mathbf{z}}$, Eq.~\eqref{eq:W_Sigma} is not well defined and we instead get
\begin{equation}
W_\Sigma(\mathbf{k}_\shortparallel)=\begin{pmatrix}
 1 & 0 & 0 & 0 \\
 0 & 0 & 0 & 1 \\
 0 & 1 & 0 & 0 \\
 0 & 0 & 1 & 0
\end{pmatrix}.
\end{equation}
This leads to the spin-up and spin-down blocks
\begin{align}
&\mathcal{H}_\sigma(\mathbf{k}) \notag \\
& = \begin{pmatrix}
\epsilon_\mathbf{k} + \sigma \lambda  l(\mathbf{k}) &
  \sigma e^{i \phi_{\mathbf{n}_{\mathbf{k}_\shortparallel}}}
  [ \Delta^s_\mathbf{k} + \sigma \Delta^t_\mathbf{k} l(\mathbf{k})] \\
  \sigma e^{-i \phi_{\mathbf{n}_{\mathbf{k}_\shortparallel}}}
  [ \Delta^s_\mathbf{k} + \sigma \Delta^t_\mathbf{k} l(\mathbf{k})] &
    -\epsilon_\mathbf{k} - \sigma \lambda  l(\mathbf{k})
\end{pmatrix} \!,
\end{align}
where $\sigma = {\uparrow}$ ($\downarrow$) is understood to have the numerical value $+1$ ($-1$) and we have introduced the phase factor
\begin{align}
e^{i \phi_{\mathbf{n}_{\mathbf{k}_\shortparallel}}}
  = \begin{dcases}\frac{n^x_{\mathbf{k}_\shortparallel}+i n^y_{\mathbf{k}_\shortparallel}}{\sqrt{(n^x_{\mathbf{k}_\shortparallel})^2
  +(n^y_{\mathbf{k}_\shortparallel})^2}}
    &\text{for } n^z_{\mathbf{k}_\shortparallel} \neq 1, \\
  1 &\text{for } n^z_{\mathbf{k}_\shortparallel} = 1.
\end{dcases}
\end{align}
The blocks $\mathcal{H}_\sigma$ do not have BdG form---in a BdG Hamiltonian, the lower right component would have to equal $-[\epsilon_{-\mathbf{k}} + \sigma \lambda l(-\mathbf{k})] = -\epsilon_\mathbf{k} + \sigma \lambda l(\mathbf{k})$ since $l(\mathbf{k})$ is odd. The blocks thus break PHS. They also break TRS separately since TRS maps $\mathcal{H}_\uparrow$ onto $\mathcal{H}_\downarrow$ and vice versa.

Each of the blocks $\mathcal{H}_\sigma$ retains a chiral symmetry, i.e., there are unitary matrices $U_{S,\sigma}(\mathbf{k}_\shortparallel)$ such that
\begin{equation}
U_{S,\sigma}^\dagger(\mathbf{k}_\shortparallel) H_\sigma(\mathbf{k}) U_{S,\sigma}(\mathbf{k}_\shortparallel) = - H_\sigma(\mathbf{k}),
\end{equation}
where $U_{S,\uparrow}(\mathbf{k}_\shortparallel)$ [$U_{S,\downarrow}(\mathbf{k}_\shortparallel)$] is the upper (lower) diagonal $2 \times 2$ block of $\tilde{U}_S(\mathbf{k}_\shortparallel)=W_\Sigma(\mathbf{k}_\shortparallel) U_S W_\Sigma^\dagger(\mathbf{k}_\shortparallel)$, i.e.,
\begin{equation}
U_{S,\sigma}(\mathbf{k}_\shortparallel) = \sigma \begin{dcases} \begin{pmatrix}
 0 & i e^{i \phi_{\mathbf{n}_{\mathbf{k}_\shortparallel}}} \\
 -i e^{-i \phi_{\mathbf{n}_{\mathbf{k}_\shortparallel}}} & 0
\end{pmatrix} &  \text{for }   n^z_{\mathbf{k}_\shortparallel} \neq 1, \\
\begin{pmatrix}
 0 & i \\
 -i & 0
\end{pmatrix} & \text{for }  n^z_{\mathbf{k}_\shortparallel} = 1.
\end{dcases}
\end{equation}
Thus each of the two blocks $\mathcal{H}_\sigma$ is in class AIII~\cite{SRFL08,AZ97,Z96,SBT12}.

\subsection{Winding number}
\label{subsec:winding}

The well-established way \cite{SR11, STY11,SBT12,SB15} to define a $\mathbf{k}_\shortparallel$-de\-pen\-dent one-dimensional winding number is to bring the Hamiltonian $\mathcal{H}(\mathbf{k})$ into block-off-diagonal form via the transformation
\begin{equation}
W_S \mathcal{H}(\mathbf{k}) W_S^\dagger=\begin{pmatrix}
0& D(\mathbf{k}) \\ D^\dagger(\mathbf{k}) &0
\end{pmatrix} \equiv \overline{\mathcal{H}}(\mathbf{k}),
\end{equation}
where $W_S$ diagonalizes $U_S$ such that
\begin{equation}
W_S U_S W_S^\dagger = \sigma_z \otimes \sigma_0.
\end{equation}
One can now adiabatically deform $\overline{\mathcal{H}}(\mathbf{k})$ into a flat-band Hamiltonian, which amounts to replacing the diagonal matrix $\Sigma_D(\mathbf{k})$ in the singular-value decomposition $D(\mathbf{k})=U_D^\dagger(\mathbf{k} ) \Sigma_D(\mathbf{k}) V_D(\mathbf{k})$ by the unit matrix. The resulting off-diagonal block $q_D(\mathbf{k}) = U_D^\dagger(\mathbf{k}) V_D(\mathbf{k})$ can be used to define a winding number
\begin{equation} \label{eq:W_D}
W_D(\mathbf{k}_\shortparallel) = \frac{1}{2 \pi i} \int_{k_\perp} dk_\perp\, \text{Tr}[ q_D^\dagger(\mathbf{k}) \partial_{k_\perp} q_D(\mathbf{k})].
\end{equation}
As the matrices $U_D^\dagger(\mathbf{k} )$ and $V_D(\mathbf{k} )$ and thus also  $ q_D(\mathbf{k})$ are unitary, the winding number can be transformed to
\begin{align}
&W_D(\mathbf{k}_\shortparallel) = \frac{1}{2 \pi i} \int_{k_\perp} dk_\perp\, \partial_{k_\perp} [\ln \det q_D(\mathbf{k})] \notag \\
&\quad = \frac{1}{2\pi i} \int_{k_\perp} dk_\perp\, \partial_{k_\perp}\big({}\ln |{\det}\, q_D(\mathbf{k})| + i \arg[\det q_D(\mathbf{k})]\big) \notag \\
&\quad = \frac{1}{2\pi} \int_{k_\perp} dk_\perp\, \partial_{k_\perp} \arg[\det q_D(\mathbf{k})].
\end{align}
One can now use
\begin{align}
&\arg[\det q_D(\mathbf{k})] = \arg[\det U_D^\dagger(\mathbf{k}) \det V_D(\mathbf{k})]
  \notag \\
&\quad = \arg[\det U_D^\dagger(\mathbf{k}) \det \Sigma_D(\mathbf{k}) \det V_D(\mathbf{k})/ {\det}\, \Sigma_D(\mathbf{k})]\notag\\
&\quad = \arg[\det D(\mathbf{k})]- \arg[\det \Sigma_D(\mathbf{k})] \notag\\
&\quad = \arg[\det D(\mathbf{k})],
\end{align}
where the contribution of $\arg[\det \Sigma_D(\mathbf{k})]$ drops out if all singular values of $D(\mathbf{k})$ are nonzero. Points $\mathbf{k}_\shortparallel$ where $D(\mathbf{k})$ has at least one singular value equal to zero coincide with points where $\det D(\mathbf{k})=0$, i.e., gap nodes. At such $\mathbf{k}_\shortparallel$, the winding number is ill defined. For all other $\mathbf{k}_\shortparallel$, we can write
\begin{equation}
W_D(\mathbf{k}_\shortparallel)= \frac{1}{2\pi} \int_{k_\perp} dk_\perp\, \partial_{k_\perp}
  \arg [\det D(\mathbf{k})],
\end{equation}
i.e., the winding number describes how many times the complex function $\det D(k_\perp,\mathbf{k}_\shortparallel)$ winds around the origin when $k_\perp$ traverses the Brillouin zone once.

We now review the case of noncentrosymmetric superconductors. For such systems, the winding number $W_D(\mathbf{k}_\shortparallel)$ can only be nonzero if the triplet-to-singlet ratio is sufficiently large to ensure that nodal lines occur on one of the helicity Fermi surfaces \cite{SBT12}. The dispersion relations of the positive-helicity and negative-helicity bands, $\xi^{\pm}(\mathbf{k})=\epsilon_\mathbf{k} \pm \lambda |\mathbf{l}_\mathbf{k}|$, are given by the two eigenvalues of the normal-state Hamiltonian $h(\mathbf{k})$. Note that the decomposition into the two helicity bands is different from the block diagonalization performed in Sec.\ \ref{sub:symmetries} since $l(\mathbf{k}) \neq |\mathbf{l}_\mathbf{k}|$. We recall that $l(\mathbf{k})$ is odd in $\mathbf{k}$, whereas $|\mathbf{l}_\mathbf{k}|$ is even.

The gaps on the corresponding two helicity Fermi surfaces are $\Delta^\pm(\mathbf{k})=\Delta^s_\mathbf{k} \pm \Delta^t_\mathbf{k} |\mathbf{l}_\mathbf{k}|$.
In these systems, $W_D(\mathbf{k}_\shortparallel)$ is nonzero for surface momenta $\mathbf{k}_\shortparallel$ that lie inside the projection of the nodal lines into the sBZ. As noted in Sec.\ \ref{sec:intro}, there is always a region outside of the projection of the nodal lines that is topologically trivial and does not support zero-energy surface states. In particular, the winding number at the origin $\mathbf{k}_\shortparallel=0$ always vanishes.

A system that avoids this problem would therefore have to break TRS while still preserving chiral symmetry, i.e., the anticommutation of the Hamiltonian with a unitary matrix of which half the eigenvalues are $+1$ and the other half are $-1$. The usual BdG Hamiltonian in Eq.~\eqref{eq:bdg_hamiltonian} cannot satisfy both conditions because its construction requires it to have PHS. However, we have shown in Sec.\ \ref{sub:symmetries} that it is possible to construct models with TRS and PHS that decompose into blocks $\mathcal{H}_\sigma(\mathbf{k})$ that break both symmetries but retain chiral symmetry. It remains to show that such models allow for a nonzero winding number.

To this end, we define a winding number analogous to $W_D(\mathbf{k}_\shortparallel)$ in Eq.~\eqref{eq:W_D} for $\mathcal{H}_{\sigma}(\mathbf{k})$, i.e.,
\begin{equation}
W_{\perp,\sigma}(\mathbf{k}_\shortparallel) = \frac{1}{2\pi} \int_{k_\perp} dk_\perp \partial_{k_ \perp} \arg[D_{\sigma}(\mathbf{k})] ,
\end{equation}
where
\begin{equation}\label{eq:D_updown}
D_{\sigma}(\mathbf{k})
  =  -\lbrace\epsilon_\mathbf{k} + \sigma \lambda l(\mathbf{k})
    + i\, [\Delta^s_\mathbf{k} + \sigma \Delta^t_\mathbf{k} l(\mathbf{k})]\rbrace
\end{equation}
are the off-diagonal entries of the matrices
\begin{align}
\overline{\mathcal{H}}_{\sigma}(\mathbf{k})&=\begin{pmatrix}
0& D_{\sigma}(\mathbf{k}) \\ D_{\sigma}^\ast(\mathbf{k}) &0
\end{pmatrix} \notag \\
&=W_{S,\sigma}(\mathbf{k}_\shortparallel) \mathcal{H}_\sigma(\mathbf{k}) W_{S,\sigma}^\dagger(\mathbf{k}_\shortparallel)
\end{align}
and
\begin{equation}
W_{S,\sigma}(\mathbf{k}_\shortparallel)=\frac{1}{\sqrt{2}}\begin{dcases}
\begin{pmatrix}
 \frac{ -i\sigma (n^x_{\mathbf{k}_\shortparallel}-i n^y_{\mathbf{k}_\shortparallel})}{ \sqrt{(n^x_{\mathbf{k}_\shortparallel})^2+(n^y_{\mathbf{k}_\shortparallel})^2}} & 1 \\[2ex]
 \frac{i \sigma (n^x_{\mathbf{k}_\shortparallel}-i n^y_{\mathbf{k}_\shortparallel})}{ \sqrt{(n^x_{\mathbf{k}_\shortparallel})^2+(n^y_{\mathbf{k}_\shortparallel})^2}} & 1
\end{pmatrix} &  \text{for }   n^z_{\mathbf{k}_\shortparallel} \neq 1, \\
\begin{pmatrix}
- i \sigma  & 1 \\ i \sigma & 1
\end{pmatrix} &  \text{for }   n^z_{\mathbf{k}_\shortparallel} = 1
\end{dcases}
\end{equation}
diagonalizes $U_{S,\sigma}(\mathbf{k}_\shortparallel)$ such that
\begin{equation}
W_{S,\sigma}(\mathbf{k}_\shortparallel) U_{S,\sigma}(\mathbf{k}_\shortparallel) W_{S,\sigma}^\dagger(\mathbf{k}_\shortparallel) = \sigma_z.
\end{equation}
We thus have
\begin{align}\label{eq:windingupdown}
W_{\perp,\sigma}(\mathbf{k}_\shortparallel)
  &= \frac{1}{2\pi} \int_{k_\perp} dk_\perp\, \partial_{k_ \perp} \notag \\
&\quad{}\times \arg\big(\epsilon_{\mathbf{k}} + \sigma \lambda l(\mathbf{k})
  + i\, [\Delta^s_{\mathbf{k}} + \sigma \Delta^t_\mathbf{k} l(\mathbf{k})]\big).
\end{align}
We note that due to $l(\mathbf{k})$ being odd in $\mathbf{k}$ the winding numbers of the $\uparrow$ and $\downarrow$ blocks are related by
\begin{equation}
W_{\perp,\uparrow }(\mathbf{k}_\shortparallel)=-W_{\perp,\downarrow}(-\mathbf{k}_\shortparallel)
\end{equation}
for the case of $(s+p)$-wave pairing with $\Delta^s = \mathrm{const}$ and $\Delta^t = \mathrm{const}$, which we have assumed in our calculations. This result still holds true as long as the admixtures of higher-moment contributions are sufficiently small and the quasiparticle energies in the bulk, i.e, the eigenvalues of the bulk BdG Hamiltonian, do not have nodes.
It is therefore sufficient to consider only one of them, e.g., $W_{\perp,\uparrow}(\mathbf{k}_\shortparallel)$, from now on.

We now assume a sufficiently weak SOC strength $\lambda$ and a dispersion $\epsilon_{k_\perp,\mathbf{k}_\shortparallel}$ that is sufficiently flat in the $\mathbf{k}_\shortparallel$ directions such that for every $\mathbf{k}_\shortparallel$ there is exactly one positive and one negative solution $k_\perp$ for
\begin{equation}
\xi^\pm_{k_\perp,\mathbf{k}_\shortparallel}
   = \epsilon_{k_\perp,\mathbf{k}_\shortparallel} \pm \lambda |\mathbf{l}_{k_\perp,\mathbf{k}_\shortparallel}| = 0 ,
\end{equation}
This means that the Fermi surface for each helicity $\pm$ consists of a corrugated sheet with $k_\perp \in (0,\pi)$ and another sheet with $k_\perp \in (-\pi,0)$. We denote the perpendicular components of the $\mathbf{k}$ points on the positive-helicity Fermi surface, i.e., the solutions $k_\perp$ of $\xi^+_{k_\perp,\mathbf{k}_\shortparallel}=0$, by $k_{\perp}^{(1)}>0$ and $k_{\perp}^{(2)}<0$ and the solutions for the negative-helicity Fermi surface by $k_{\perp}^{(3)}>0$ and $k_{\perp}^{(4)}<0$.

The resulting conditions for nonzero winding numbers in the entire sBZ read as
\begin{align}\label{eq:gapcondition1}
\text{sgn}\Big(\Delta^+_{k_{\perp}^{(1)},\mathbf{k}_\shortparallel}\Big) &=
- \text{sgn}\Big(\Delta^-_{k_{\perp}^{(4)},\mathbf{k}_\shortparallel}\Big) , \\
\label{eq:gapcondition2}
\text{sgn}\Big(\Delta^+_{k_{\perp}^{(2)},\mathbf{k}_\shortparallel}\Big) &=
- \text{sgn}\Big(\Delta^-_{k_{\perp}^{(3)},\mathbf{k}_\shortparallel}\Big),
\end{align}
where $\Delta^{\pm}(\mathbf{k})= \Delta^s_{\mathbf{k}} \pm \Delta^t_{\mathbf{k}}|\mathbf{l}_{\mathbf{k}}|$. A detailed derivation is given in Appendix~\ref{sec:appenix_1}. Since $\Delta^+$ is always positive for $\Delta^s, \Delta^t>0$ this means that $\Delta^-$ must be negative everywhere on the negative-helicity Fermi surface, which requires a sufficiently small singlet-to-triplet ratio $\Delta^s/\Delta^t<\min |l(\mathbf{k}_{F}^-)|$.

\section{Ideal system}\label{sec:ideal_system}
\subsection{Crystallographic point groups}

In this section, we determine for which point groups one can find parameter regimes where the winding number $W_{\perp,\sigma}(\mathbf{k}_\shortparallel)$ in Eq.~\eqref{eq:windingupdown} is nonzero for all momenta $\mathbf{k}_\shortparallel$ in the sBZ for some surface orientation.
As shown in Sec.~\ref{subsec:winding}, this requires the matrices $\mathcal{H}(\mathbf{k})$ and $\Sigma_{\mathbf{n}_{\mathbf{k}_\shortparallel}}$ to commute. This is equivalent to the requirement that a unit vector $\mathbf{n}_{\mathbf{k}_\shortparallel}$ independent of $k_\perp$ exists such that the vectors $\mathbf{g}_{k_\perp,\mathbf{k}_\shortparallel}$ and $\mathbf{d}_{k_\perp,\mathbf{k}_\shortparallel}$ and hence $\mathbf{l}_{k_\perp,\mathbf{k}_\shortparallel}$ are parallel to $\mathbf{n}_{\mathbf{k}_\shortparallel}$. There is no point group that guarantees this relation but there are several point groups for which one can at least fine tune the parameters in such a way that the symmetry holds. Point groups with inversion symmetry can be excluded since they are incompatible with a nonzero SOC vector $\mathbf{l}_{k_\perp,\mathbf{k}_\shortparallel}$. Of the 32 crystallographic point groups, the 21 groups $C_1$, $C_2$, $C_3$, $C_4$, $C_6$, $D_2$, $D_3$, $D_4$, $D_6$, $C_{2v}$, $C_{3v}$, $C_{4v}$, $C_{6v}$, $D_{2d}$, $C_s$, $C_{3h}$, $D_{3h}$, $S_4$, $T$, $O$, and $T_d$ are noncentrosymmetric. $T$, $O$, and $T_d$, cannot satisfy the symmetry in Eq.~\eqref{eq:spin_symmetry}, even approximately. This can be seen from the fact that, for any surface orientation, Eq.~\eqref{eq:group_relation_l} requires the existence of a vector $\mathbf{n}_{\mathbf{k}_\shortparallel}$ for every point $\mathbf{k}_\shortparallel$ in the corresponding sBZ such that
\begin{equation} \label{eq:condition_n_point_group}
\mathbf{n}_{\mathbf{k}_\shortparallel}= \alpha_{\mathbf{k}_\shortparallel} R\, \mathbf{n}_{R^{-1}\mathbf{k}_\shortparallel}
\end{equation}
for some real number $\alpha_{\mathbf{k}_\shortparallel}$ and every element $R$ of the point group. However, all the cubic groups contain the four $C_3$ axes of a regular tetrahedron. For a threefold rotation, Eq.~\eqref{eq:condition_n_point_group} can only be satisfied if the vector $\mathbf{n}_{\mathbf{k}_\shortparallel}$ is parallel to the axis of rotation. It would thus have to be parallel to all the four $C_3$ axes at once, which is impossible.

For the other noncentrosymmetric point groups, those including a three-, four-, or sixfold rotation axis, i.e., $C_3$, $C_4$, $C_6$, $D_3$, $D_4$, $D_6$, $C_{3v}$, $C_{4v}$, $C_{6v}$, $C_{3h}$, and $D_{3h}$, as well as $S_4$, which contains a fourfold rotation-reflection axis, and $D_{2d}$, which has three orthogonal $C_2$ axes and two mirror planes containing one of the axes, the same equation leads to the restriction that the SOC vector must be parallel to the principal axis, which we will define as the $z$-axis. For the groups $D_2$ and $C_{2v}$, Eq.~\eqref{eq:condition_n_point_group} would allow for the SOC vector to be parallel to any of the three coordinate axes $x$, $y$, or $z$. Continuity then forces it to be unidirectional in the entire sBZ. For the point groups $C_2$, which contains only one twofold rotation axis $z$, and $C_s$, which contains the mirror plane $xy$, the SOC vector may either be parallel to the $z$-axis or have an arbitrary orientation within the $xy$ plane. The $C_1$ point group does not imply any restrictions on the orientation of $\mathbf{n}_{\mathbf{k}_\shortparallel}$. Hence, only the groups $C_2$, $C_s$, and $C_1$ allow for the orientation of the SOC vector to vary as a function of $\mathbf{k}_\shortparallel$. In this work, we assume a unidirectional SOC vector and leave potential effects due to its $\mathbf{k}_\shortparallel$ for $C_2$, $C_s$, and $C_1$ for the future.

To find out which of these point groups are the most promising to obtain a unidirectional SOC vector and therefore have a nonzero winding number in the entire sBZ for some surface orientation, we consider the lowest-order expansion of the SOC vector. Taking the periodicity of the lattice into account, this can be done by writing
\begin{equation}\label{eq:lattice_expansion_SOC}
\mathbf{l}_\mathbf{k}=\sum_{j,k,l}\mathbf{c}_{j,k,l}\,
  \sin[\mathbf{k}\cdot (j \mathbf{g}_1+k \mathbf{g}_2+l \mathbf{g}_3)] ,
\end{equation}
with lattice vectors $\mathbf{g}_1$, $\mathbf{g}_2$, $\mathbf{g}_3$ and vector-valued coefficients $\mathbf{c}_{j,k,l}$ that respect the lattice symmetries \cite{S09,SSYA17}. We expect the dominant contribution in $\mathbf{l_k}$ to come from nearest-neighbor terms. For a nonzero winding number, we not only need to write the SOC vector as $\mathbf{l_k}=l(\mathbf{k}) \mathbf{n}_{\mathbf{k}_\shortparallel}$ but we also require $l(\mathbf{k})$ not to be even in $k_\perp$. If $l(\mathbf{k})$ were even in $k_\perp$ the image of the function $D_\sigma(\mathbf{k})$ in Eq.~\eqref{eq:D_updown} as a function of $k_\perp$ would be a curve in the complex plane that does not enclose any area and therefore cannot lead to a nonzero winding number. It is therefore necessary that $l(\mathbf{k})$ is not even in $k_\perp$.

Equation \eqref{eq:lattice_expansion_SOC} shows that the point groups $C_2$, $C_3$, $C_4$, $C_6$, $D_2$, $D_3$, $D_4$, and $D_6$ lead to a unidirectional SOC vector with $k_\perp=k_z$ if only the coefficients of terms parallel to $\hat{\mathbf{z}}$ are nonzero. This corresponds to a SOC vector proportional to $\sin k_z\; \hat{\mathbf{z}}$ to first order. For the point group $D_2$, one can alternatively also set $k_\perp=k_x$ if the SOC vector is proportional to $\sin k_x\; \hat{\mathbf{x}}$ or $k_\perp=k_y$ if the SOC vector is proportional to $\sin k_y\; \hat{\mathbf{y}}$ to first order.
For the point groups $C_2$ and $C_s$, the three surface orientations $(100)$, $(010)$, and $(001)$, i.e., $k_\perp\in\lbrace k_x, k_y, k_z \rbrace$ can all lead to a first-order approximation $l(\mathbf{k})=\sin k_\perp$ and a nonzero winding number in the full sBZ. In these cases, the SOC vector is oriented normal to the corresponding surface. In principle, higher-index surfaces containing the $z$-axis are also permitted since the SOC vector may point in any direction in the $xy$ plane. However, this leads to backfolding of the sBZ and is thus incompatible with the special shape of the normal-state Fermi surface assumed in our model.

For the point group $C_{2v}$, the most general form of the SOC vector on an orthorhombic lattice is
\begin{align}\label{eq:SOC_C2v}
\mathbf{l_k}&=\sum_{j,k,l\, =\, 0}^{\infty}\begin{pmatrix} c_{j,k,l}^{x} \cos(j k_x) \sin(k k_y)  \cos(l k_z) \\ c_{j,k,l}^{y} \sin(j k_x) \cos(k k_y)  \cos(l k_z) \\ c_{j,k,l}^{z} \sin(j k_x) \sin(k k_y)  \sin(l k_z) \end{pmatrix} \notag \\
&= (c_{0,1,0}^x \sin k_y+\dots)\, \hat{\mathbf{x}}
  + (c_{1,0,0}^y \sin k_x+\dots)\, \hat{\mathbf{y}} \notag \\
&\quad{}+ (c_{1,1,1}^z \sin k_x \sin k_y \sin k_z +\dots)\, \hat{\mathbf{z}},
\end{align}
with real coefficients $c_{j,k,l}^{x}$, $c_{j,k,l}^{y}$, and $c_{j,k,l}^{z}$.
This can be derived from Eq.~\eqref{eq:lattice_expansion_SOC} by expanding $\sin(j k_x + k k_y + l k_z)$ and using the symmetries to derive conditions for the coefficients $\mathbf{c}_{j,k,l}$. For example, for the $z$-component of the SOC vector, the two-fold rotation symmetry leads to $l^z_{k_x, k_y, k_z}=l^z_{-k_x, -k_y, k_z}$, from which we can conclude $c^z_{j,k,l}=c^z_{-j,-k,l}$, and the two mirror planes lead to $l^z_{k_x, k_y, k_z}=l^z_{k_x, -k_y, k_z}$ ($xz$-plane) and $l^z_{k_x, k_y, k_z}=l^z_{-k_x, k_y, k_z}$ ($yz$-plane), i.e., $c^z_{j,k,l}=c^z_{j,-k,l}=c^z_{-j,k,l}$. Combining these conditions, rewriting the sum in Eq.~\eqref{eq:lattice_expansion_SOC} such that $j$, $k$, and $l$ are positive, and redefining $c^z_{j,k,l}$ to absorb all constant prefactors leads to the $z$-component in Eq.~(\ref{eq:SOC_C2v}). We therefore again find the options $k_\perp \in \lbrace k_{x},k_y \rbrace$ for
\begin{align}
\mathbf{l_k}&= \sin k_{x} [(c_{1,0,0}^{y} +c_{1,1,0}^{y} \cos k_{y} +c_{1,0,1}^{y} \cos k_z \notag \\&\quad{} + c_{1,1,1}^{y} \cos k_{y} \cos k_z )\hat{\mathbf{y}}+\sin k_{y} \sin k_z \hat{\mathbf{z}}]
\end{align} and
\begin{align}
\mathbf{l_k}&= \sin k_{y} [(c_{1,0,0}^{x} +c_{1,1,0}^{x} \cos k_{x} +c_{1,0,1}^{x} \cos k_z \notag \\&\quad{} + c_{1,1,1}^{x} \cos k_{x} \cos k_z )\hat{\mathbf{x}}+\sin k_{x} \sin k_z \hat{\mathbf{z}}]
\end{align}
to lowest order. Focusing only on nearest-neighbor terms, this leads to $\mathbf{l_k}\propto \sin k_{x} \hat{\mathbf{y}}$ for $k_\perp=k_x$ and $\mathbf{l_k}\propto \sin k_{y} \hat{\mathbf{x}}$ for $k_\perp=k_y$. For this point group, the choice $\mathbf{n}_{\mathbf{k}_\shortparallel} = \hat{\mathbf{z}}$ cannot lead to $W_{\perp,\sigma}(\mathbf{k}_\shortparallel)\neq 0$ for all $\mathbf{k}_\shortparallel$, because even if the prefactors of $\sin k_y\; \hat{\mathbf{x}}$ and $\sin k_x\; \hat{\mathbf{y}}$ are arbitrarily small compared to the coefficients $c_{j,k,l}^z$ the SOC vector rotates from an orientation parallel to $\hat{\mathbf{z}}$ to one parallel to $\hat{\mathbf{y}}$ or $\hat{\mathbf{x}}$ in a sufficiently small neighborhood of $k_x\rightarrow 0$ or $k_y\rightarrow 0$, respectively, so that the spin symmetry is broken there.

An analogous problem occurs for $\mathbf{n}_{\mathbf{k}_\shortparallel} = \hat{\mathbf{z}}$ for the point groups $C_{3v}$, $C_{4v}$, $C_{6v}$, $D_{2d}$, $C_{3h}$, $D_{3h}$, and $S_4$. All of these groups include more than just twofold rotation symmetries or a single mirror plane so that the $z$-axis is the only orientation of a unidirectional SOC vector that is not forbidden by Eq.~\eqref{eq:condition_n_point_group}. However, like for $C_{2v}$, $\mathbf{n}_{\mathbf{k}_\shortparallel} = \hat{\mathbf{z}}$ cannot lead to a nonzero winding number in the full sBZ. Therefore, one cannot get a nonzero winding number in the whole sBZ for any surface orientation for these point groups.

This leaves the eleven point groups $C_1$, $C_2$, $C_3$, $C_4$, $C_6$, $D_2$, $D_3$, $D_4$, $D_6$, $C_{2v}$, and $C_s$ which can all lead to a nonzero winding number in the whole sBZ if the parameters in the SOC vector are fine-tuned appropriately. A detailed list of the possible options can be found in Table~\ref{tab:SOC_general} in Appendix~\ref{sec:appenix_2}. Superconductors belonging to several of these point groups exist, e.g., Ir$_2$Ga$_9$ \cite{WAKKMIAA09}, Rh$_2$Ga$_9$ \cite{WAKKMIAA09}, and Y$_3$Pt$_4$Ge$_{13}$ \cite{GNASSTLG13} in $C_s$, BiPd \cite{JTR11, MJKKGTMRR12, MMSMTJRKAZ13} and UIr under pressure \cite{AHFKYHSO04, AHKKFYHSO04, KFHKAYHSO06} in $C_2$, LaNiC$_2$ \cite{HY09,CJZCYNSY13} and ThCoC$_2$ \cite{GMKF14} in $C_{2v}$, as well as (Ta,Nb)Rh$_2$B$_2$ \cite{CXK18, MOF21} with $C_3$. To our knowledge, the triplet-to-singlet pairing ratio, which needs to be large for our scenario, is unknown for these compounds. The possible exception is (Ta,Nb)Rh$_2$B$_2$, which might have line nodes \cite{MOF21}, suggesting a sizable triplet component. However, the presence of line nodes of course precludes our scenario.

\subsection{Results}\label{subsec:results}

\begin{figure}[!htbp]
\includegraphics[width=0.8\columnwidth,trim={0 0.1cm 0 0.25cm},clip]{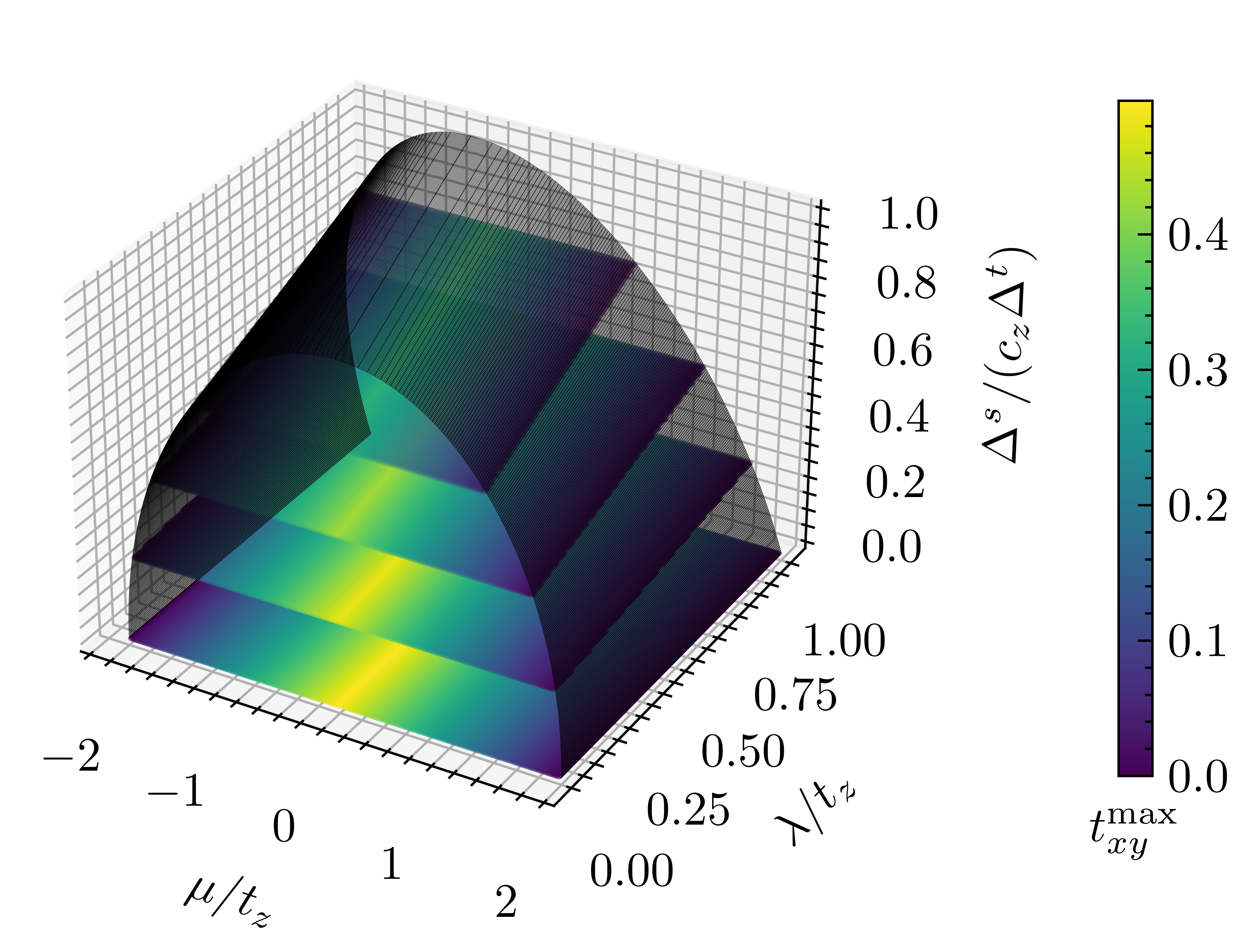}
\caption{Parameter regime which ensures a nonzero winding number in the entire sBZ for the model of Sec.~\ref{subsec:results}. The boundary of the region in the three-dimensional space $(\tilde{\mu}, \tilde{\lambda}, \Delta^s/(c_z \Delta^t))$ where this is possible is shown in semi-transparent gray. The colored planes show the maximal possible value of $\tilde{t}_{xy}$ such that the winding number is still nonzero everywhere in the sBZ.}
\label{fig:winding_region}
\end{figure}

For the eleven point groups $C_1$, $C_2$, $C_3$, $C_4$, $C_6$, $D_2$, $D_3$, $D_4$, $D_6$, $C_{2v}$, and $C_s$, there is at least one surface orientation $k_\perp\in\lbrace k_x, k_y, k_z \rbrace$  such that the nearest-neighbor approximation for the SOC vector is
\begin{equation}
\mathbf{l_k} = \sin k_\perp\: \mathbf{n},
\end{equation}
with a unit vector $\mathbf{n} = \mathbf{n}_{\mathbf{k}_\shortparallel}$.

In this section, we will assume that the parameters are chosen such that there is a vector $\mathbf{n}$ which is exactly parallel to $\mathbf{l}_{k_\perp,\mathbf{k}_\shortparallel}$ for all $\mathbf{k}$. The experimentally more realistic case, in which this condition is only approximately satisfied, is discussed in Sec.~\ref{sec:nonideal_systems}.
In order to get a nonzero winding number, it is also necessary that the gaps on the positive-helicity and negative-helicity Fermi surfaces have opposite signs. The gap $\Delta^+_{\mathbf{k}_F^+}$ on the positive-helicity Fermi surface always has positive sign if we assume $\Delta^s_\mathbf{k}=\Delta^s>0$ and $\Delta^t_\mathbf{k}=\Delta^t>0$ so that this is equivalent to requiring the gap $\Delta^-_{\mathbf{k}_F^-}$ to be negative. Here, we use the definition of the Fermi momenta $\mathbf{k}_F^\pm=(k_{\perp}^{(i)}, \mathbf{k}_\shortparallel)$ from Sec.~\ref{subsec:winding}, where they were given as the solutions $\mathbf{k}$ of $\xi_\mathbf{k}^\pm=\epsilon_\mathbf{k} \pm \lambda |\mathbf{l}_\mathbf{k}|=0$. This means that the projection of the Fermi surfaces has to cover the entire sBZ in order to get solutions $k_{\perp}^{(i)}$ for all $\mathbf{k}_\shortparallel$ in the sBZ. We assume a SOC vector ${\mathbf{l}_{k_\perp,\mathbf{k}_\shortparallel} \parallel \mathbf{n}}$ with $l({\mathbf{k}})=c_z \sin k_z$ and $k_\perp=k_z$. For the dispersion $\epsilon_\mathbf{k}$, we choose a model with $C_4$ point-group symmetry on a tetragonal lattice. We assume a nearest-neighbor hopping amplitude $t_z$ in the $z$ direction and a hopping amplitude $t_{xy}$ in the $x$ and $y$ directions as well as a chemical potential $\mu$, i.e.,
\begin{equation}\label{eq:eps_model}
\epsilon_\mathbf{k}=-\mu-2 t_z \cos k_z-2 t_{xy} (\cos k_x +\cos k_y).
\end{equation}
Changing the dispersion, e.g., to account for a different point group, does not qualitatively alter the following results, and calculations for different directions of $\mathbf{n}$ and $k_\perp$ can be done analogously and also lead to similar results.

For every point $\mathbf{k}_\shortparallel$ in the sBZ, the two solutions $k_\perp^{(3),(4)}$ of the equation $\epsilon_{\mathbf{k}}-\lambda |l(k_z)|=0$, i.e., the negative-helicity Fermi momenta, can be calculated, yielding
\begin{widetext}
\begin{equation}
 k_\perp^{(3),(4)}= 2 \arctan \frac{\tilde{\lambda}  \pm  \sqrt{
    4  + \tilde{\lambda}^2
    -\left[-\tilde{\mu} -2 \tilde{t}_{xy} \left(\cos k_x+\cos k_y\right)\right]^2}}{2 + \left[-\tilde{\mu} -2 \tilde{t}_{xy} \left(\cos k_x+\cos k_y\right)\right]}
\end{equation}
if $4+\tilde{\lambda} ^2 -\left[\tilde{\mu} + 2 \tilde{t}_{xy} \left(\cos k_x+\cos k_y\right) \right]^2>0$, where we have introduced the dimensionless parameters $\tilde{t}_{xy}=t_{xy}/t_z$, $\tilde{\mu}=\mu/t_z$, and $\tilde{\lambda}=c_z \lambda/t_z$.
We therefore get a nonzero winding number if $l_\text{min}< \Delta^s/(c_z \Delta^t) < l_\text{max}$ with
\begin{align}
l_\text{min} &= \max\left( 0, \max_{k_x,k_y} \frac{\tilde{\lambda} \left[-\tilde{\mu} -2 \tilde{t}_{xy} \left(\cos k_x+\cos k_y\right)\right] - 2 \sqrt{4 + \tilde{\lambda}^2 - \left[-\tilde{\mu} -2 \tilde{t}_{xy} \left(\cos k_x+\cos k_y\right)\right]^2 }}{4 + \tilde{\lambda}^2} \right), \\
\label{eq:l_max}
l_\text{max} &= \min_{k_x,k_y} \frac{\tilde{\lambda} \left[-\tilde{\mu} -2 \tilde{t}_{xy} \left(\cos k_x+\cos k_y\right)\right] + 2 \sqrt{4 + \tilde{\lambda}^2 - \left[-\tilde{\mu} -2 \tilde{t}_{xy} \left(\cos k_x+\cos k_y\right)\right]^2 }}{4 + \tilde{\lambda}^2}.
\end{align}
\end{widetext}

This means that the winding number is nonzero in the entire sBZ for a subset of nonzero measure of the parameter space $(\tilde{\mu}, \tilde{\lambda}, \Delta^s/(c_z \Delta^t), \tilde{t}_{xy})$. Figure~\ref{fig:winding_region} shows the boundary of the projection of this region into the three-dimensional parameter space $(\tilde{\mu}, \tilde{\lambda}, \Delta^s/(c_z \Delta^t))$ in gray. The colored planes in this region indicate below which maximal value of $\tilde{t}_{xy}$ a nonzero winding number at that point $(\tilde{\mu}, \tilde{\lambda}, \Delta^s/(c_z \Delta^t))$ is still ensured in the whole sBZ. Figure~\ref{fig:winding_region} shows that the dispersion $\epsilon_\mathbf{k}$ must be sufficiently flat in the $\mathbf{k}_\shortparallel$ direction, i.e., $\tilde{t}_{xy}$ must be sufficiently small, and that $\Delta^s/(c_z \Delta^t)$ must be sufficiently small, i.e., the spin-triplet component of the pairing must be sufficiently strong compared to the spin-singlet pairing.

\begin{figure*}[!htbp]
\centering
\includegraphics[width=16.256cm]{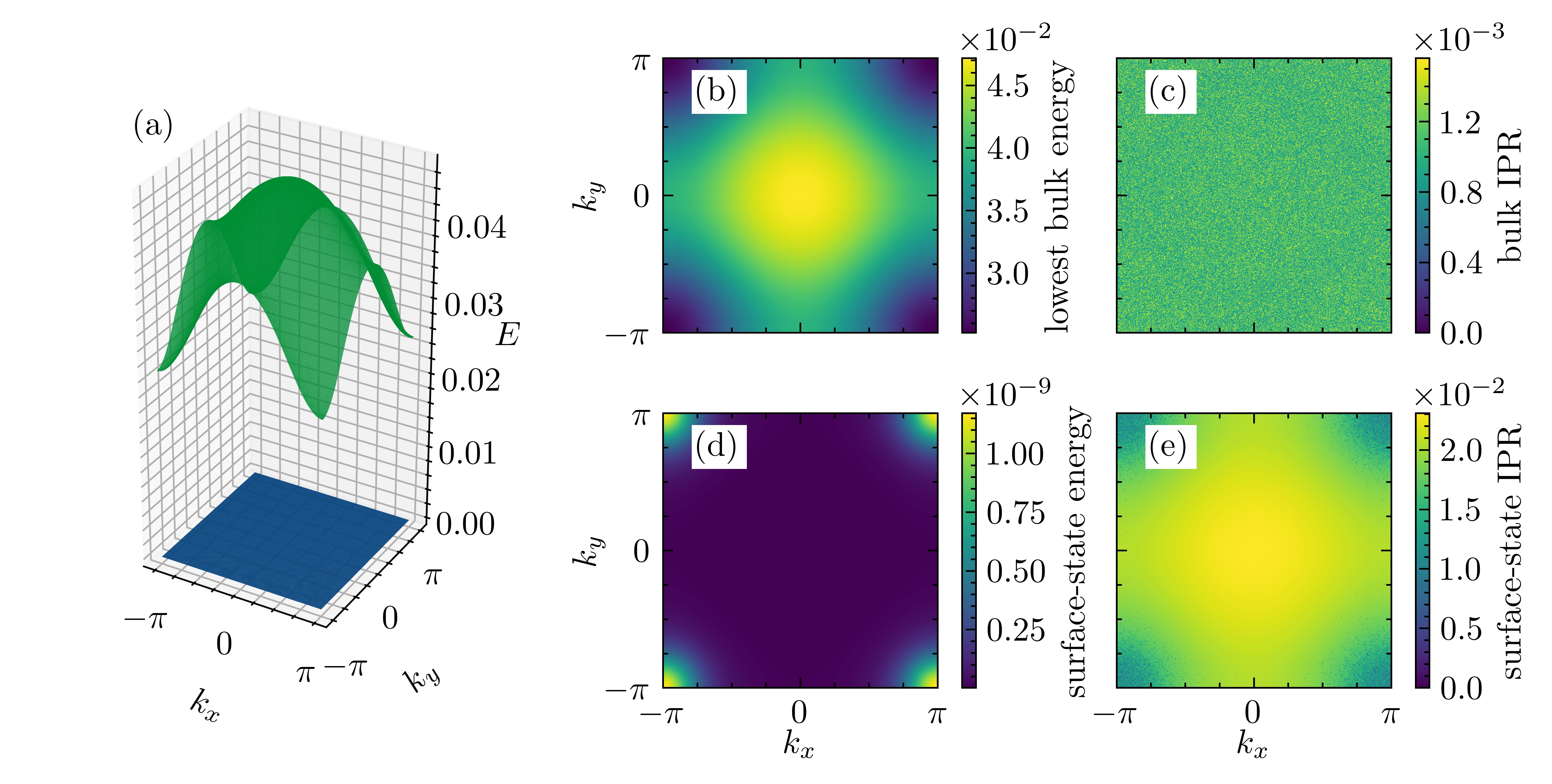}
\caption{Energy and IPR of the surface state and the bulk state with the lowest energy at given $\mathbf{k}_\shortparallel = (k_x, k_y)$ in an ideal system with nonzero winding number $W_{\perp,\sigma}(\mathbf{k}_\shortparallel)$ in the full sBZ. (a) Surface-state energy and lowest bulk energy, illustrating the size of the gap. (b) Lowest bulk energy as a density plot. (c) IPR of the corresponding state, where low IPR corresponds to a delocalized state and high IPR corresponds to a strongly localized state. (d) Energy and (e) IPR of the surface state. Note that in panels (b) to (e) the same color scheme is used for different orders of magnitude.}
\label{fig:C4_ideal}
\end{figure*}

As an example, \Cref{fig:C4_ideal} shows the energy and inverse participation ratio (IPR) of the surface state and of the bulk state with the lowest energy at given $\mathbf{k}_\shortparallel = (k_x, k_y)$ for a system with $t_z=1$, $t_{xy}=0.1$, $\mu=-1$, $\lambda=0.1$, $\Delta^s=0.1$, $\Delta^t=0.2$, and $c_z=1$ calculated by exact diagonalization of the BdG Hamiltonian on a slab with $Z=500$ layers. The derivation of the specific matrix that has to be diagonalized is presented in Appendix~\ref{sec:appendix_3}. The IPR
\begin{equation}
I[\Psi(z_i, k_\shortparallel)]=\sum_{z_i}|\Psi(z_i, k_\shortparallel)|^4 ,
\end{equation}
where $\Psi$ is the normalized wave function, measures the localization of a state, i.e., a localized surface state has a higher IPR than a delocalized bulk state.\ Figure \ref{fig:C4_ideal}(a) shows the energies of both states, which illustrates that the bulk is indeed fully gapped.
Here, the bulk state with the lowest energy refers to the eigenstate of the slab Hamiltonian at given $\mathbf{k}_\shortparallel=(k_x,k_y)$ that has the minimal positive eigenenergy among those states that are not localized at the surface, as inferred from the IPR. In Figs.\ \ref{fig:C4_ideal}(b) and \ref{fig:C4_ideal}(c), we plot the energy of the bulk state and the corresponding IPR, respectively. Figure \ref{fig:C4_ideal}(d) shows the energy of the surface state, which is an almost flat band very close to zero energy. The IPR of the surface state is plotted in Fig.\ \ref{fig:C4_ideal}(e). When comparing the IPR of the surface state and of the bulk state, note that the same color scheme was used for different orders of magnitude. It is obvious from the comparison of Figs.\ \ref{fig:C4_ideal}(c) and \ref{fig:C4_ideal}(e) that the IPR of the surface state is much higher than the IPR of the bulk state because the former is localized at the surfaces of the slab, while the latter is spread out over all layers.

\begin{figure}[!htbp]
\includegraphics[width=0.8\columnwidth]{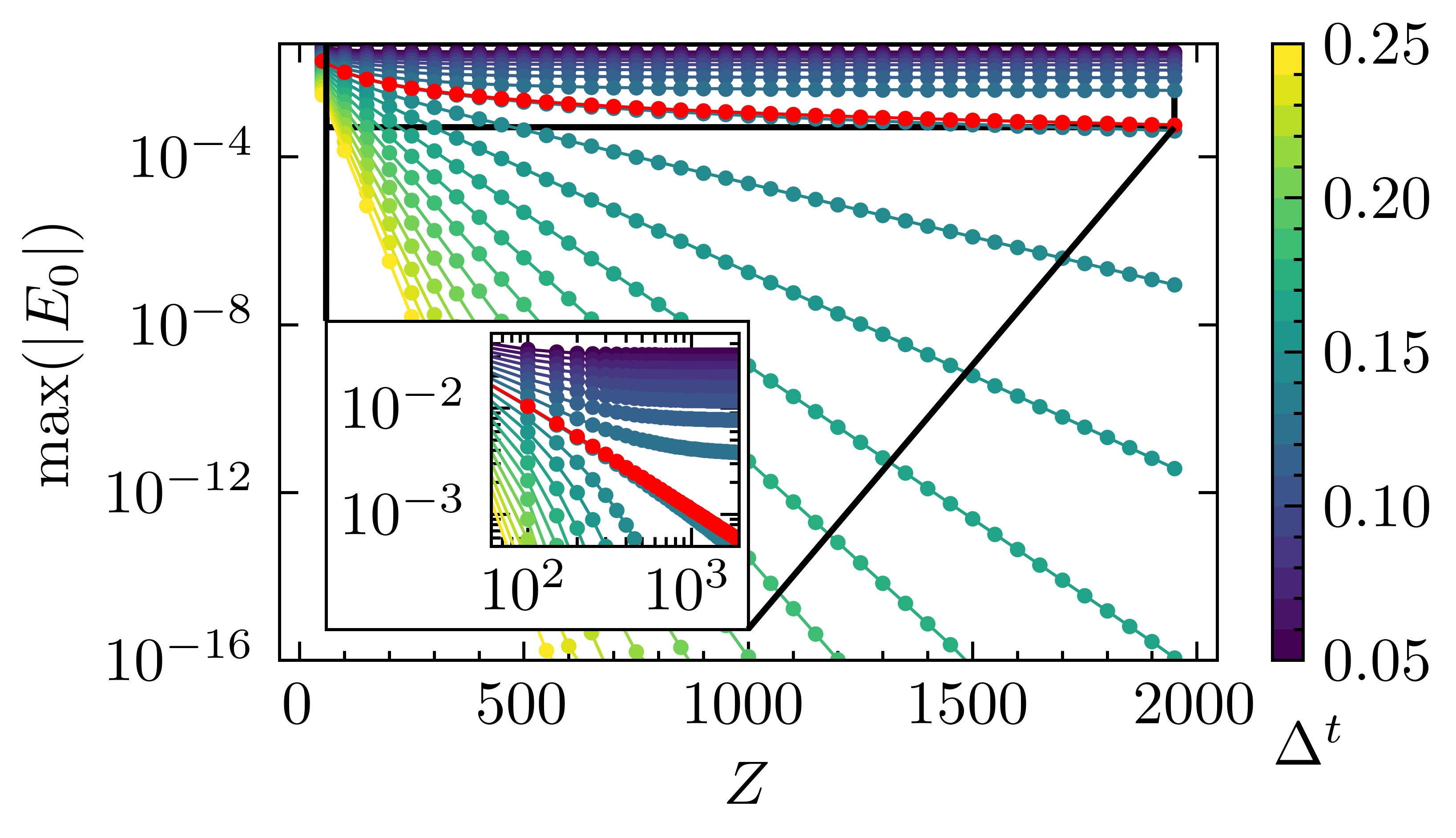}
\caption{Lowest eigenvalue of the Hamiltonian as a function of the thickness $Z$ for various values of the triplet pairing amplitude $\Delta^t$. For sufficiently large values of $\Delta^t$, the winding number ensures the existence of a zero-energy surface state in the limit of an infinitely thick slab, while for a slab of finite thickness, the energy of the surface state decreases exponentially. If $\Delta^t$ is too small, there are no zero-energy surface states and the energy approaches a finite value. The inset shows the same data for small $\Delta^t$ on a double logarithmic scale. The red line indicates the transition between zero-energy surface states and states at finite energy.}\label{fig:thicknessdependence_E}
\end{figure}

Figure~\ref{fig:thicknessdependence_E} demonstrates that for sufficiently small $\Delta^s/\Delta^t$, the energy of the surface state indeed decreases exponentially with increasing thickness $Z$. The finite energy of this state within the parameter region with a nonzero winding number is therefore a consequence of the finite thickness. Outside of this parameter region, the energy approaches a constant nonzero value for $Z \rightarrow \infty$. This transition from nonzero to zero winding number can be seen in the inset, where the former corresponds to the lines with negative curvature while the latter corresponds to lines with positive curvature.
The transition is marked by the red line in Fig.\ \ref{fig:thicknessdependence_E}, which has been determined by calculating $\Delta^t = \Delta^s/(c_z l_{\max})$ where $l_{\max}$ is given by Eq.~\eqref{eq:l_max}. This leads to a value of $\Delta^t\approx 0.1335$, which has been used to obtain the red line.

\section{Non-ideal systems} \label{sec:nonideal_systems}

In this section, we consider terms in the Hamiltonian that break the symmetries described in Sec.~\ref{sec:ideal_system}. By construction, PHS is always present in the BdG formalism. Therefore, chiral symmetry and time-reversal symmetry are equivalent. These symmetries could for example be broken by introducing an exchange field at the surface of the slab \cite{BTS13, STB13, LT22}, which would couple to the spin polarization of the surface states. In the case where the spin-rotation symmetry is still preserved, i.e., the field points along the spin axis, this leads to a shift of the flat bands in energy. The shift depends on the (very small) range of the exchange field and on the decay length of the surface state into the bulk, i.e., on its localization at the surface. If the exchange field breaks not only TRS but also spin-rotation symmetry, i.e., if the field is not parallel to the spin-rotation-symmetry axis, the leading energy correction still comes from the field component that couples to the spin polarization of the unperturbed system. Only if the field is orthogonal to the spin rotation axis the leading term in perturbation theory is of quadratic order in the field.

However, in real physical systems, the idealized situation described in Sec.~\ref{sec:ideal_system} is unlikely to occur even in the absence of an applied field because the spin symmetry in Eq.~\eqref{eq:spin_symmetry} can only be achieved by fine tuning parameters in the SOC vector. In this section, we therefore focus on the case where this symmetry is broken: It is plausible to assume that while one can find systems where the SOC vector will point in approximately the same direction for varying momentum $k_\perp$ perpendicular to the surface, there will generically be small deviations from this direction. If the spin symmetry is not present, the winding number $W_{\perp,\sigma}(\mathbf{k}_\shortparallel)$ cannot be defined and there are no flat bands of zero-energy surface states. However, we can still consider the BdG Hamiltonian
\begin{equation}
\mathcal{H}(\mathbf{k})=\mathcal{H}^{(0)}(\mathbf{k})+\mathcal{H}^{(1)}(\mathbf{k})
\end{equation}
as the sum of the unperturbed Hamiltonian
\begin{equation}
\mathcal{H}^{(0)}(\mathbf{k})=\begin{pmatrix}
\epsilon_\mathbf{k} \sigma_0+\lambda \mathbf{l}^{\shortparallel}_\mathbf{k} {\cdot} \boldsymbol{\sigma} & \!\!\!(\Delta^s_\mathbf{k} + \Delta^t_\mathbf{k} \mathbf{l}^{\shortparallel}_\mathbf{k} {\cdot} \boldsymbol{\sigma})(i \sigma_y) \\
-(i \sigma_y)(\Delta^s_\mathbf{k} + \Delta^t_\mathbf{k} \mathbf{l}^{\shortparallel}_\mathbf{k} {\cdot} \boldsymbol{\sigma})& -\epsilon_\mathbf{k} \sigma_0 + \lambda \mathbf{l}^{\shortparallel}_\mathbf{k} {\cdot} \boldsymbol{\sigma}^\ast
\end{pmatrix} \!,
\end{equation}
which only contains the part $\mathbf{l}^\shortparallel_\mathbf{k} \equiv (\mathbf{l}_\mathbf{k}\cdot \mathbf{n}_{\mathbf{k}_\shortparallel})\mathbf{n}_{\mathbf{k}_\shortparallel}\equiv l^\shortparallel(\mathbf{k}) \mathbf{n}_{\mathbf{k}_\shortparallel}$ of the SOC vector parallel to $\mathbf{n}_{\mathbf{k}_\shortparallel}$ and a small perturbation
\begin{equation}
\mathcal{H}^{(1)}(\mathbf{k})=\begin{pmatrix}
\lambda \mathbf{l}^{\perp}_\mathbf{k} \cdot \boldsymbol{\sigma} & \Delta^t_\mathbf{k} (\mathbf{l}^{\perp}_\mathbf{k} \cdot \boldsymbol{\sigma})(i \sigma_y) \\
- \Delta^t_\mathbf{k} (i \sigma_y)(\mathbf{l}^{\perp }_\mathbf{k} \cdot \boldsymbol{\sigma})& \lambda \mathbf{l}^{\perp }_\mathbf{k} \cdot \boldsymbol{\sigma}^\ast
\end{pmatrix},
\end{equation}
which contains the components $\mathbf{l}^{\perp}_\mathbf{k}=\mathbf{l}_\mathbf{k}-(\mathbf{l}_\mathbf{k}\cdot \mathbf{n}_{\mathbf{k}_\shortparallel})\mathbf{n}_{\mathbf{k}_\shortparallel}$ orthogonal to $\mathbf{n}_{\mathbf{k}_\shortparallel}$. In this section, we will use degenerate first-order perturbation theory to estimate the energy of the surface states of $\mathcal{H}(\mathbf{k})$, employing an approximation for the surface states of $\mathcal{H}^{(0)}(\mathbf{k})$ which we obtain by a quasiclassical approximation in the direction orthogonal to the surface.

The normal-state block of the unperturbed Hamiltonian has the eigenvalues $\xi^\pm_\mathbf{k}= \epsilon_\mathbf{k}  \pm \lambda |\mathbf{l}^\shortparallel_{\mathbf{k}}| = \epsilon_\mathbf{k}  \pm \lambda | l^\shortparallel(\mathbf{k})|$ with the corresponding eigenvectors
\begin{align} \label{eq:eigvecs}
\mathbf{v}^\pm_\mathbf{k}=\begin{dcases}
\left(1,\pm \frac{n^x_{\mathbf{k}_\shortparallel} + i n^y_{\mathbf{k}_\shortparallel}}{\text{sgn}[ l^\shortparallel(\mathbf{k})]\pm n^z_{\mathbf{k}_\shortparallel}}\right)^{\!T} \hspace{-0.8em} &  \text{for }   n^z_{\mathbf{k}_\shortparallel} \neq 1, \\
\left\lbrace (1,0)^T , (0,1)^T  \right\rbrace &  \text{for } n^z_{\mathbf{k}_\shortparallel} = 1,  l^\shortparallel(\mathbf{k})>0,\\
\left\lbrace (0,1)^T, (1,0)^T  \right\rbrace&  \text{for } n^z_{\mathbf{k}_\shortparallel} = 1,  l^\shortparallel(\mathbf{k})<0.
\end{dcases}
\end{align}
As the matrix $\Delta^s_\mathbf{k} + \Delta^t_\mathbf{k} \mathbf{l}^{\shortparallel}_\mathbf{k} \cdot \boldsymbol{\sigma}$ commutes with the normal-state block, the two matrices are simultaneously diagonalizable, which means that the vectors in Eq.~\eqref{eq:eigvecs} are also eigenvectors of $\Delta^s_\mathbf{k} + \Delta^t_\mathbf{k} \mathbf{l}^{\shortparallel}_\mathbf{k} \cdot \boldsymbol{\sigma}$. The corresponding eigenvalues are $\Delta^\pm_\mathbf{k}=\Delta^s_\mathbf{k} \pm \Delta^t_\mathbf{k} |l^\shortparallel(\mathbf{k})|$.

We assume that the unperturbed Hamiltonian is fully gapped, thus we expect the gap to also stay open in the perturbed system as long as the perturbation is sufficiently small.
We also know that there is a finite region of the parameter space where $W_{\perp,\uparrow}(\mathbf{k}_\shortparallel)=W_{\perp,\uparrow}=1$ for all momenta in the sBZ and that this leads to $|W_{\perp,\uparrow}|+|W_{\perp,\downarrow}|=2\, |W_{\perp,\uparrow}|=2$ protected zero-energy surface states in the unperturbed system.

For every point $\mathbf{k}_\shortparallel$ in the sBZ, there are four points on the positive-helicity and negative-helicity Fermi surfaces that are projected onto this point. As in Sec.\ \ref{subsec:winding}, the solutions of $\xi^+_{k_\perp,\mathbf{k}_\shortparallel}=0$ are denoted by $k_{\perp}^{(1)}>0$ and $k_{\perp}^{(2)}<0$, while the solutions of $\xi^-_{k_\perp,\mathbf{k}_\shortparallel}=0$ are denoted by $k_{\perp}^{(3)}>0$ and $k_{\perp}^{(4)}<0$. We make the ansatz~\cite{BST15}
\begin{align}
&\Psi(r_\perp, k_\shortparallel)
= a_{\mathbf{k}_\shortparallel}^{(1)}\, \Psi^+(k_{\perp}^{(1)}, \mathbf{k}_\shortparallel)
  \exp\Big(i k_{\perp}^{(1)} r_\perp
  - \kappa^{+}_{k_{\perp}^{(1)},\mathbf{k}_\shortparallel} r_\perp\Big) \notag \\
&\quad{} + a_{\mathbf{k}_\shortparallel}^{(2)}\, \Psi^+(k_{\perp}^{(2)}, \mathbf{k}_\shortparallel)
  \exp\Big(i k_{\perp}^{(2)} r_\perp -\kappa^{+}_{k_{\perp}^{(2)},\mathbf{k}_\shortparallel}  r_\perp\Big) \notag \\
&\quad{} + a_{\mathbf{k}_\shortparallel}^{(3)}\, \Psi^-(k_{\perp}^{(3)}, \mathbf{k}_\shortparallel)
  \exp\Big(i k_{\perp}^{(3)} r_\perp -\kappa^{-}_{k_{\perp}^{(3)},\mathbf{k}_\shortparallel}  r_\perp\Big) \notag \\
&\quad{} + a_{\mathbf{k}_\shortparallel}^{(4)}\, \Psi^-(k_{\perp}^{(4)}, \mathbf{k}_\shortparallel)
  \exp\Big(i k_{\perp}^{(4)} r_\perp -\kappa^{-}_{k_{\perp}^{(4)},\mathbf{k}_\shortparallel} r_\perp\Big)
\end{align}
for the wave function of the surface state, where $\kappa^{\pm}_{\mathbf{k}}$ is the inverse decay length
\begin{align}
\label{eq:inversedecaylength}
\kappa_\mathbf{k}^\pm=\frac{\sqrt{|\Delta^\pm_\mathbf{k}|^2-E^2}}{\hbar |v_{\perp,F}^\pm|} \stackrel{E=0}{=}\frac{|\Delta^\pm_\mathbf{k}|}{\hbar |v_{\perp,F}^\pm|},
\end{align}
with the Fermi velocity perpendicular to the surface,
\begin{equation}
v_{\perp,F}^\pm = \frac{1}{\hbar}\, \frac{\partial \xi^\pm_\mathbf{k}}
  {\partial k_\perp}\bigg|_{\mathbf{k}=(k_\perp^F,\mathbf{k}_\shortparallel)} .
\end{equation}
The spinors $\Psi^\pm$ are defined as
\begin{align}
\Psi^{\pm} (k_{\perp}^\pm, \mathbf{k}_\shortparallel)
  &= \sqrt{\displaystyle\frac{\text{sgn}[l^\shortparallel(\mathbf{k})] \pm n^z_{\mathbf{k}_\shortparallel}}
    {4\, \text{sgn}[ l^\shortparallel(\mathbf{k})]}}\:
    \left(1, \pm n,\mp \gamma^\pm_{\mathbf{k}} n, \gamma^\pm_{\mathbf{k}} \right)^T ,
\end{align}
 with $n=(n^x_{\mathbf{k}_\shortparallel} + i n^y_{\mathbf{k}_\shortparallel})/(\text{sgn}[ l^\shortparallel(\mathbf{k})]\pm n^z_{\mathbf{k}_\shortparallel})$, which is only well defined for $n^z_{\mathbf{k}_\shortparallel} \neq 1$. Moreover, $k_\perp^\pm$ is chosen such, that $\mathbf{k}\equiv(k_\perp^\pm, \mathbf{k}_\shortparallel)=\mathbf{k}_F$ is a Fermi momentum, i.e., $k_\perp^+\in\lbrace k_{\perp}^{(1)}, k_{\perp}^{(2)}\rbrace$  and $k_\perp^-\in\lbrace k_{\perp}^{(3)}, k_{\perp}^{(4)}\rbrace$. For $\mathbf{n}_{\mathbf{k}_\shortparallel}=\hat{\mathbf{z}}$, we instead find
\begin{equation}
\Psi^{\pm}(k_{\perp}^\pm,\mathbf{k}_\shortparallel) = \left\lbrace
  \frac{1}{\sqrt{2}} \left(1,0,0,\gamma^+_{\mathbf{k}}\right)^T,
  \frac{1}{\sqrt{2}} \left(0,1,-\gamma^-_{\mathbf{k}},0\right)^T \right\rbrace
\end{equation}
for $l^\shortparallel(\mathbf{k})>0$ and
\begin{equation}
\Psi^{\pm}(k_{\perp}^\pm,\mathbf{k}_\shortparallel) = \left\lbrace
  \frac{1}{\sqrt{2}} \left(0,1,-\gamma^+_{\mathbf{k}},0\right)^T,
  \frac{1}{\sqrt{2}} \left(1,0,0,\gamma^-_{\mathbf{k}}\right)^T \right\rbrace
\end{equation}
for $l^\shortparallel(\mathbf{k})<0$.
Moreover, we have defined
\begin{equation}
\gamma^\pm_{\mathbf{k}}
  = \frac{1}{\Delta^\pm_\mathbf{k}}
  \left[E-i~\text{sgn}(v_{\perp,F}^\pm)\sqrt{|\Delta^\pm_\mathbf{k}|^2-E^2}\right] ,
\end{equation}
which for $E=0$ becomes
\begin{align}
\gamma^\pm_{\mathbf{k}} &= -i~\text{sgn}(v_{\perp,F}^\pm)~\text{sgn}(\Delta^\pm_\mathbf{k})
  \notag \\
&= \begin{dcases}
  -i& \text{for } k_{\perp}^+=k_{\perp}^{(1)} \text{ and for } k_\perp^-=k_{\perp}^{(4)}, \\
  i& \text{for } k_{\perp}^+=k_{\perp}^{(2)} \text{ and for } k_\perp^-=k_{\perp}^{(3)}.
\end{dcases}
\end{align}
Our approach relies on the quasiclassical assumption that the system is continuous in the direction perpendicular to the surface, which is located at $r_\perp=0$. For a surface state, the coefficients $a^{(i)}$ have to be chosen such that the wave function vanishes at the surface, i.e., $\Psi(r_\perp{=}0,\mathbf{k}_\shortparallel)=0$ and is normalized, i.e.,
\begin{align}
\int_{0}^\infty dr_\perp \Psi^\dagger(r_\perp,\mathbf{k}_\shortparallel) \Psi(r_\perp,\mathbf{k}_\shortparallel)=1.
\end{align}
The first condition means that the coefficient vector $(a_{\mathbf{k}_\shortparallel}^{(1)},a_{\mathbf{k}_\shortparallel}^{(2)},a_{\mathbf{k}_\shortparallel}^{(3)},a_{\mathbf{k}_\shortparallel}^{(4)})^T$ must be in the kernel $\text{ker}(\mathcal{M})=\lbrace (-1,0,0,1)^T,(0,-1,1,0)^T \rbrace$ of the matrix
\begin{align}
\mathcal{M} &= \left(\Psi^{+}\big(\mathbf{k}^{(1)}\big),
  \Psi^{+}\big(\mathbf{k}^{(2)}\big),
  \Psi^{-}\big(\mathbf{k}^{(3)}\big),
  \Psi^{-}\big(\mathbf{k}^{(4)}\big) \right) ,
\end{align}
where $\mathbf{k}^{(i)} = (k_{\perp}^{(i)},\mathbf{k}_\shortparallel)$.
Therefore, there is a two-di\-men\-sion\-al eigenspace of zero-energy surface states for $\mathcal{H}^{(0)}$, which is spanned by the vectors $\Psi_1(r_\perp,\mathbf{k}_\shortparallel)$ and $\Psi_2(r_\perp,\mathbf{k}_\shortparallel)$ with
\begin{widetext}
\begin{align}
&\Psi_1(r_\perp,\mathbf{k}_\shortparallel) = a^{(1)}_{\mathbf{k}_\shortparallel}
  \left[-\exp\left({i k_{\perp}^{(1)} r_\perp -\kappa_{\mathbf{k}_F}^{(1)} r_\perp}\right)
  + \exp\left({i k_{\perp}^{(4)} r_\perp -\kappa_{\mathbf{k}_F}^{(4)} r_\perp}\right)\right]
  \notag \\
&\qquad{} \times \begin{dcases}
  \sqrt{\displaystyle
    \frac{\text{sgn}[l^\shortparallel(k_{\perp}^{(1)},\mathbf{k}_\shortparallel)]
     + n^z_{\mathbf{k}_\shortparallel}}
  {4\, \text{sgn}[l^\shortparallel(k_{\perp}^{(1)},\mathbf{k}_\shortparallel)]}}
  \left(1, \frac{n^x_{\mathbf{k}_\shortparallel}
  + i n^y_{\mathbf{k}_\shortparallel}}
  {\text{sgn}[ l^\shortparallel(k_{\perp}^{(1)},\mathbf{k}_\shortparallel)]
  + n^z_{\mathbf{k}_\shortparallel}},i \frac{n^x_{\mathbf{k}_\shortparallel}
  + i n^y_{\mathbf{k}_\shortparallel}}
  {\text{sgn}[ l^\shortparallel(k_{\perp}^{(1)},\mathbf{k}_\shortparallel)]
  + n^z_{\mathbf{k}_\shortparallel}}, -i \right)^{\!\!T}
  &\text{for } n^z_{\mathbf{k}_\shortparallel} \neq 1, \\
  \frac{1}{\sqrt{2}} \left(1,0,0,-i\right)^T
  &\text{for } n^z_{\mathbf{k}_\shortparallel} = 1, l^\shortparallel(\mathbf{k})>0, \\
  \frac{1}{\sqrt{2}} \left(0,1,i,0\right)^T
  &\text{for } n^z_{\mathbf{k}_\shortparallel} = 1,  l^\shortparallel(\mathbf{k})<0,
\end{dcases} \\
&\Psi_2(r_\perp,\mathbf{k}_\shortparallel)=a^{(2)}_{\mathbf{k}_\shortparallel}
  \left[-\exp\left({i k_{\perp}^{(2)} r_\perp -\kappa_{\mathbf{k}_F}^{(2)} r_\perp}\right)
  + \exp\left({i k_{\perp}^{(3)} r_\perp -\kappa_{\mathbf{k}_F}^{(3)} r_\perp}\right)\right]
  \notag \\
&\qquad{} \times \begin{dcases}
  \sqrt{\displaystyle
    \frac{\text{sgn}[l^\shortparallel(k_{\perp}^{(1)},\mathbf{k}_\shortparallel)]
     - n^z_{\mathbf{k}_\shortparallel}}
  {4 \text{sgn}[ l^\shortparallel(k_{\perp}^{(1)},\mathbf{k}_\shortparallel)]}}
  \left(1, -\frac{n^x_{\mathbf{k}_\shortparallel}
  + i n^y_{\mathbf{k}_\shortparallel}}
  {\text{sgn}[l^\shortparallel(k_{\perp}^{(1)},\mathbf{k}_\shortparallel)]
  - n^z_{\mathbf{k}_\shortparallel}},i \frac{n^x_{\mathbf{k}_\shortparallel}
  + i n^y_{\mathbf{k}_\shortparallel}}
  {\text{sgn}[ l^\shortparallel(k_{\perp}^{(1)},\mathbf{k}_\shortparallel)]
  - n^z_{\mathbf{k}_\shortparallel}}, i \right)^{\!\!T}
  &\text{for } n^z_{\mathbf{k}_\shortparallel} \neq 1, \\
  \frac{1}{\sqrt{2}} \left(0,1,-i,0\right)^T
  &\text{for } n^z_{\mathbf{k}_\shortparallel} = 1,  l^\shortparallel(\mathbf{k})>0, \\
  \frac{1}{\sqrt{2}}\left(1,0,0,i\right)^T
  &\text{for } n^z_{\mathbf{k}_\shortparallel} = 1,  l^\shortparallel(\mathbf{k})<0,
\end{dcases}
\end{align}
 where we have used $\text{sgn}[l^\shortparallel( k_{\perp}^{(1)},\mathbf{k}_\shortparallel)]=\text{sgn}[l^\shortparallel(k_{\perp}^{(3)},\mathbf{k}_\shortparallel)]=-\text{sgn}[l^\shortparallel(k_{\perp}^{(2)},\mathbf{k}_\shortparallel)]=-\text{sgn}[l^\shortparallel(k_{\perp}^{(4)},\mathbf{k}_\shortparallel)]$. To ensure normalization, the coefficients $a_{\mathbf{k}_\shortparallel}^{(i)}$ have to be chosen as
\begin{align}
 a^{(1)}_{\mathbf{k}_\shortparallel}&=\left[\int_0^\infty dr_\perp \left| -\exp\left({i  k_{\perp}^{(1)} r_\perp -\kappa_{\mathbf{k}_F}^{(1)} r_\perp}\right)+\exp\left({i k_{\perp}^{(4)} r_\perp -\kappa_{\mathbf{k}_F}^{(4)} r_\perp}\right) \right|^2\right]^{-1/2}\notag \\
 &=\left[\int_0^\infty dr_\perp\, 2 e^{-\left(\kappa_{\mathbf{k}_F}^{(1)}+\kappa_{\mathbf{k}_F}^{(4)}\right) r_\perp} \left[\cosh\left(\kappa_{\mathbf{k}_F}^{(1)} r_\perp - \kappa_{\mathbf{k}_F}^{(4)} r_\perp\right) -\cos\left(k_{\perp}^{(1)} r_\perp - k_{\perp}^{(4)} r_\perp\right)\right] \right]^{-1/2},\\
 a^{(2)}_{\mathbf{k}_\shortparallel}&=\left[\int_0^\infty dr_\perp \left|-\exp\left({i k_{\perp}^{(2)} r_\perp -\kappa_{\mathbf{k}_F}^{(2)} r_\perp}\right)+\exp\left({i k_{\perp}^{(3)} r_\perp -\kappa_{\mathbf{k}_F}^{(3)} r_\perp}\right)\right|^2\right]^{-1/2}\notag \\
 &=\left[\int_0^\infty dr_\perp 2 e^{-\left(\kappa_{\mathbf{k}_F}^{(2)}+\kappa_{\mathbf{k}_F}^{(3)}\right) r_\perp} \left[\cosh\left(\kappa_{\mathbf{k}_F}^{(2)} r_\perp - \kappa_{\mathbf{k}_F}^{(3)} r_\perp\right) -\cos\left(k_{\perp}^{(2)} r_\perp - k_{\perp}^{(3)} r_\perp\right)\right] \right]^{-1/2}.
\end{align}
\end{widetext}
The first-order corrections to the energy of the surface state for the full Hamiltonian $\mathcal{H}=\mathcal{H}^{(0)}+\mathcal{H}^{(1)}$ can be calculated by diagonalizing the $2 \times 2$ matrix $\mathcal{P} $ with entries
\begin{align}
\mathcal{P}_{ij}&=\langle \Psi_i |\mathcal{H}^{(1)}| \Psi_j\rangle \notag\\
&=\int_0^\infty dr_\perp \Psi_i^\dagger(r_\perp,\mathbf{k}_\shortparallel) \mathcal{H}^{(1)}(-i \partial_{r_\perp},\mathbf{k}_\shortparallel) \Psi_j(r_\perp,\mathbf{k}_\shortparallel),
\end{align}
where the first argument of $\mathcal{H}^{(1)}(\mathbf{k})=\mathcal{H}^{(1)}(k_\perp,\mathbf{k}_\shortparallel)$ is replaced by $-i\partial_{k_\perp}$ (note $\hbar=1$), acting on the wavefunction $\Psi_j(r_\perp,\mathbf{k}_\shortparallel)$. If we assume that only the lowest-order terms in $\mathbf{l}^\perp_\mathbf{k}$ are relevant, i.e., that $\mathbf{l}^\perp_\mathbf{k}$ does not depend on $k_\perp$, we get the first-order energy corrections
\begin{align}
E^{(1)}_\pm(\mathbf{k}_\shortparallel)
  &= \pm a^{(1)}_{\mathbf{k}_\shortparallel} a^{(2)}_{\mathbf{k}_\shortparallel}\sqrt{(\Delta^t)^2+\lambda^2} ~|\mathbf{l}_\mathbf{k}^\perp|
  \int_0^\infty dr_\perp
  \notag\\
&\quad{} \times
  \left( e^{-i k_{\perp}^{(1)} r_\perp -\kappa_{\mathbf{k}_F}^{(1)} r_\perp}
  - e^{-i k_{\perp}^{(4)} r_\perp -\kappa_{\mathbf{k}_F}^{(4)} r_\perp} \right)
  \notag \\
&\quad{} \times \left( e^{i k_{\perp}^{(3)} r_\perp
  - \kappa_{\mathbf{k}_F}^{(3)} r_\perp}-e^{i k_{\perp}^{(2)} r_\perp
  - \kappa_{\mathbf{k}_F}^{(2)} r_\perp} \right)
\end{align}
of the eigenvalue corresponding to the surface state. We thus find
\begin{equation}\label{eq:E1correction}
E^{(1)}_\pm(\mathbf{k}_\shortparallel)
  = \pm \alpha_{\mathbf{k}_\shortparallel} \sqrt{(\Delta^t)^2+\lambda^2}\:
    |\mathbf{l}_\mathbf{k}^\perp|
\end{equation}
with the prefactor
\begin{widetext}
\begin{align}
\alpha_{\mathbf{k}_\shortparallel} &= 2 \left[\left(\kappa_{\mathbf{k}_F}^{(1)}
  + \kappa_{\mathbf{k}_F}^{(2)}+\kappa_{\mathbf{k}_F}^{(3)}+\kappa_{\mathbf{k}_F}^{(4)}\right)^2
  + \left(k_{\perp}^{(1)}-k_{\perp}^{(2)}-k_{\perp}^{(3)}+k_{\perp}^{(4)}\right)^2
  \right]^{1/2}
  \left(\frac{\kappa_{\mathbf{k}_F}^{(1)} \kappa_{\mathbf{k}_F}^{(2)} \kappa_{\mathbf{k}_F}^{(3)}
  \kappa_{\mathbf{k}_F}^{(4)}}
  {\left(\kappa_{\mathbf{k}_F}^{(1)}+\kappa_{\mathbf{k}_F}^{(4)}\right)
  \left(\kappa_{\mathbf{k}_F}^{(2)}+\kappa_{\mathbf{k}_F}^{(3)}\right)}\right)^{\!1/2} \notag \\
&\quad{} \times \left(\frac{\left(\kappa_{\mathbf{k}_F}^{(1)}+\kappa_{\mathbf{k}_F}^{(4)}\right)^2
  + \left(k_{\perp}^{(1)}-k_{\perp}^{(4)}\right)^2}
  {\left[\left(\kappa_{\mathbf{k}_F}^{(1)}+\kappa_{\mathbf{k}_F}^{(2)}\right)^2
  + \left(k_{\perp}^{(1)}-k_{\perp}^{(2)}\right)^2\right] \left[\left(\kappa_{\mathbf{k}_F}^{(1)}+\kappa_{\mathbf{k}_F}^{(3)}\right)^2
  + \left(k_{\perp}^{(1)}-k_{\perp}^{(3)}\right)^2\right]}\right)^{\!\!1/2} \notag \\
&\quad{} \times \left(\frac{\left(\kappa_{\mathbf{k}_F}^{(2)}+\kappa_{\mathbf{k}_F}^{(3)}\right)^2
  + \left(k_{\perp}^{(2)}-k_{\perp}^{(3)}\right)^2}
  {\left[\left(\kappa_{\mathbf{k}_F}^{(2)}+\kappa_{\mathbf{k}_F}^{(4)}\right)^2
  + \left(k_{\perp}^{(2)}-k_{\perp}^{(4)}\right)^2\right] \left[\left(\kappa_{\mathbf{k}_F}^{(3)}+\kappa_{\mathbf{k}_F}^{(4)}\right)^2
  + \left(k_{\perp}^{(3)}-k_{\perp}^{(4)}\right)^2\right]}\right)^{\!\!1/2}.
\end{align}
\end{widetext}
For practical purposes, it is useful to compare the energy $E^{(1)}_\pm(\mathbf{k}_\shortparallel)$ to the energy of the lowest-lying bulk state, i.e., with the gap size. The minimal gap size is $E_\text{bulk,min}=\min_{\mathbf{k}_\shortparallel}|\Delta^-_{\mathbf{k}_F}|$, such that
\begin{equation}
\frac{E^{(1)}_\pm(\mathbf{k}_\shortparallel)}{E_\text{bulk,min}}
  = \pm \alpha_{\mathbf{k}_\shortparallel} \frac{\sqrt{(\Delta^t)^2+\lambda^2}\:
    |\mathbf{l}_\mathbf{k}^\perp|}
    {\Delta^t \min_{\mathbf{k}_\shortparallel}
    |\mathbf{l}_{\mathbf{k}_F}|-\Delta^s}.
\end{equation}
\begin{figure*}[!htbp]
\centering
\includegraphics[width=16.256cm]{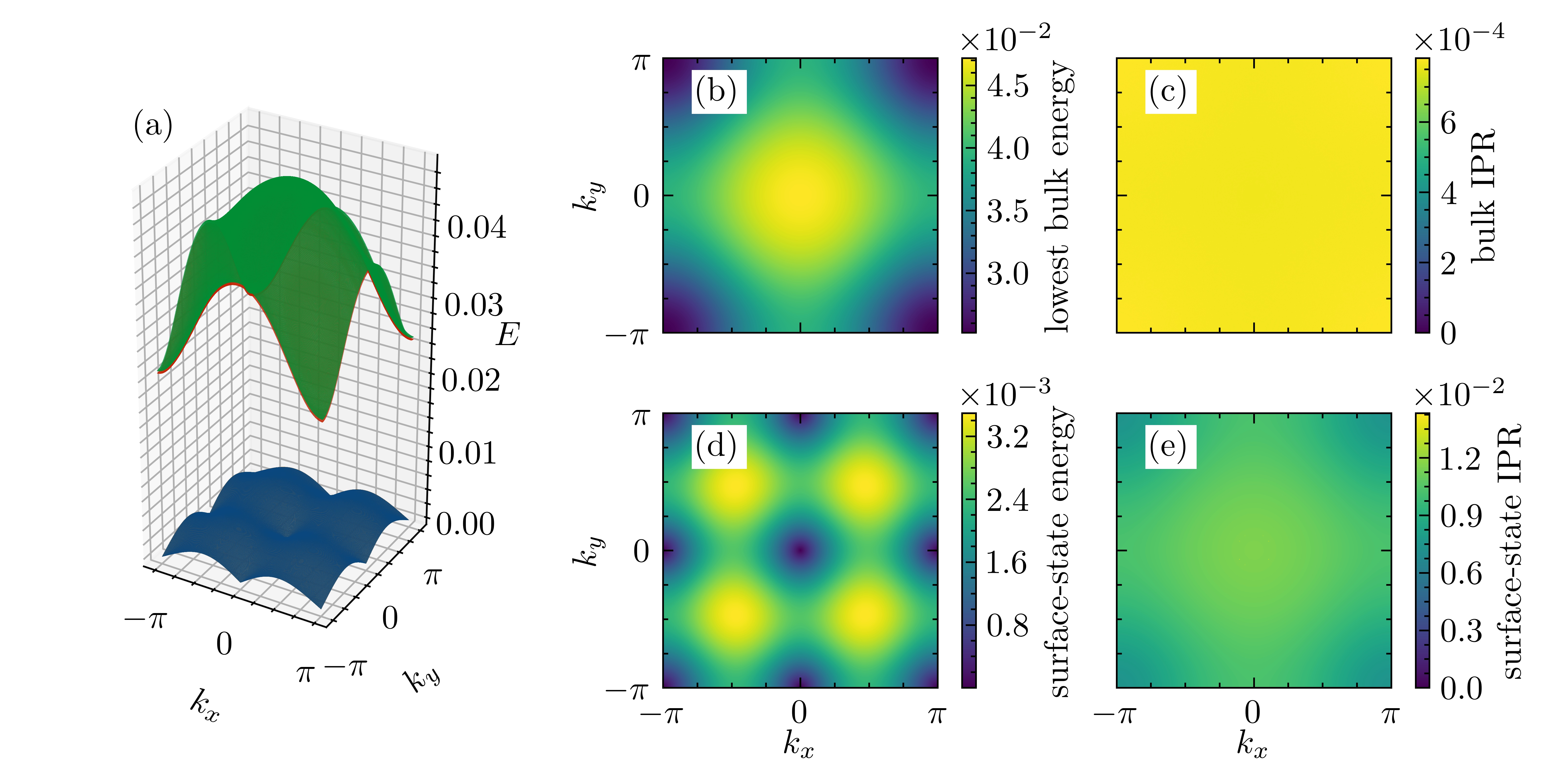}
\caption{Energy and IPR of the surface state and the bulk state with the lowest energy at given $\mathbf{k}_\shortparallel = (k_x, k_y)$ for the $C_4$ point group and a SOC vector with nonzero components $\mathbf{l}_{\mathbf{k}}^\shortparallel$ with $c_{1,0,0}^x=c_{0,1,0}^x=0.025,c_{0,0,1}^z=1$. (a) Surface-state energy from first-order perturbation theory (blue) and from exact diagonalization (orange, obscured by and indistinguishable from the blue surface) and lowest bulk energy from exact diagonalization (green) and from the minimal gap (red, nearly obscured by the green surface), illustrating the size of the gap. (b) Lowest bulk energy as a density plot. (c) IPR of the corresponding state, where low IPR corresponds to a delocalized state and high IPR corresponds to a strongly localized state. (d) Energy and (e) IPR of the surface state.}
\label{fig:C4_nonideal}
\end{figure*}

We now simplify the model further, in order to be able to calculate $\alpha_{\mathbf{k}_\shortparallel}$ analytically. To this end, we assume the momentum dependence of the SOC vector of the unperturbed system  to be $l^\shortparallel(\mathbf{k})=\sin k_\perp$ and the dispersion $\epsilon_{k_\perp, \mathbf{k}_\shortparallel}=\epsilon_{-k_\perp, \mathbf{k}_\shortparallel}$ to be even in $k_\perp$. The first assumption is reasonable for most of the noncentrosymmetric point groups which can exhibit a nonzero winding number $W_{\perp, \sigma}(\mathbf{k}_\shortparallel)$ in the full sBZ, if we assume that the SOC vector is dominated by nearest-neighbor terms, as shown in Appendix~\ref{sec:appenix_2}. The second assumption means that hopping terms in any direction which is neither parallel nor orthogonal to the surface should be negligible. For instance, this is the case if we assume dispersions consisting only of nearest-neighbor hopping terms. These assumptions lead to $k_{\perp}^{(2)}=-k_{\perp}^{(1)}$, $k_{\perp}^{(4)}=-k_{\perp}^{(3)}$, and $\kappa_{\mathbf{k}_F}^{(1)}=\kappa_{\mathbf{k}_F}^{(2)}$, $\kappa_{\mathbf{k}_F}^{(3)}=\kappa_{\mathbf{k}_F}^{(4)}$ so that the prefactor is
\begin{align}
\alpha_{\mathbf{k}_\shortparallel} &= \frac{\kappa_{\mathbf{k}_F}^{(1)}
  \kappa_{\mathbf{k}_F}^{(3)}}{\kappa_{\mathbf{k}_F}^{(1)}+\kappa_{\mathbf{k}_F}^{(3)}}\,
  \frac{\left(\kappa_{\mathbf{k}_F}^{(1)}+\kappa_{\mathbf{k}_F}^{(3)}\right)^2
  + \left(k_{\perp}^{(1)}+k_{\perp}^{(3)}\right)^2}
  {\left(\kappa_{\mathbf{k}_F}^{(1)}+\kappa_{\mathbf{k}_F}^{(3)}\right)^2
  + \left(k_{\perp}^{(1)}-k_{\perp}^{(3)}\right)^2} \notag \\
&\quad{} \times \frac{1}
  {\sqrt{\left(\kappa_{\mathbf{k}_F}^{(1)}\right)^2+\left(k_{\perp}^{(1)}\right)^2}
  \sqrt{\left(\kappa_{\mathbf{k}_F}^{(3)}\right)^2+\left(k_{\perp}^{(3)}\right)^2}}.
\end{align}

We now compare Eq.\ \eqref{eq:E1correction} with the result of exact numerical diagonalization of the BdG Hamiltonian of a slab with a $(001)$ surface and $Z=500$ layers in the $z$ direction for a system with $C_4$ point-group symmetry. Other possible point groups and surface orientations are discussed and compared in Appendix~\ref{sec:appenix_2}.
The general form of the SOC vector in this point group is
\begin{equation}
\mathbf{l_k}=\sum_{j,k,l}\begin{pmatrix} c_{j,k,l}^x \sin(j k_x+ k k_y) \cos(l k_z)\\c_{j,k,l}^x \sin(j k_y- k k_x) \cos(l k_z)\\c_{j,k,l}^z \cos(j k_x+ k k_y)\cos(k k_x- j k_y) \sin(l k_z)\end{pmatrix},
\end{equation}
with the real coefficients $c_{j,k,l}^x$ and $c_{j,k,l}^z$. Restricting this to nearest-neighbor terms, we get
\begin{align}
\mathbf{l}_\mathbf{k}&=(c_{1,0,0}^x \sin k_x+c^x_{0,1,0} \sin k_y)\: \hat{\mathbf{x}} \notag \\
&\quad{} + (c_{1,0,0}^x \sin k_y-c^x_{0,1,0} \sin k_x)\: \hat{\mathbf{y}} \notag\\
&\quad{} + c_{0,0,1}^z \sin k_z\: \hat{\mathbf{z}}.
\end{align}
In order to obtain a SOC vector which is approximately parallel to the $z$-axis, we require $|c_{1,0,0}^x|, |c^x_{0,1,0}| \ll |c_{0,0,1}^z|$. Equation \eqref{eq:E1correction} leads to the approximation
\begin{align}
E^{(1)}_\pm(\mathbf{k}_\shortparallel)
  &= \alpha_{\mathbf{k}_\shortparallel} \sqrt{2}\,
  \sqrt{(\Delta^t)^2+\lambda^2}\:
  \sqrt{(c_{1,0,0}^x)^2+(c^x_{0,1,0})^2} \notag \\
&\quad{} \times \sqrt{\sin^2 k_x+\sin^2 k_y}
\end{align}
for the surface-state energy. This relation shows that the energy of the perturbed surface band and thus the band width are of first order in the perturbation, i.e., in the SOC terms that break the symmetry in Eq.\ (\ref{eq:spin_symmetry}).

Figure \ref{fig:C4_nonideal} shows the energy and IPR of the surface state and the first bulk state for the point group $C_4$, with the parameters for $\mathcal{H}^{(0)}$ being $t_z=1$, $t_{xy}=0.1$, $\mu=-1$, $\lambda=0.1$, $\Delta^s=0.1$, and $\Delta^t=0.2$, as in Sec.\ \ref{sec:ideal_system} and the parameters for $\mathcal{H}^{(1)}$ being $c_{1,0,0}^x=c_{0,1,0}^x=0.025$ and $c_{0,0,1}^z=1$. Figure \ref{fig:C4_nonideal}(a) shows both the energy of the surface state and the gap. For the surface-state energy, the result of the exact diagonalization of the BdG Hamiltonian is given in blue, while the result of first-order perturbation theory is plotted in orange. Both results lie almost exactly on top of each other, with the relative difference between the two never exceeding $7 \times 10^{-3}$ such that the orange and blue plot cannot be distinguished in the figure. A plot of the relative difference between the two energies is given in Appendix~\ref{sec:appenix_2} in Fig.~\ref{fig:difference}. For the first bulk state, i.e., the gap size, the value from the exact diagonalization is plotted in green. The plot in red shows the gap
\begin{equation}
  \min_{k_\perp} \Delta^-_{k_\perp,\mathbf{k}_\shortparallel}
     = \Delta^-_{k_{\perp, F}^-, \mathbf{k}_\shortparallel} .
\end{equation}
Both plots should coincide in the limit of an infinitely thick slab, while for finite thickness, the result from exact diagonalization should be slightly higher, due to the discretization of momentum space. In Fig.~\ref{fig:C4_nonideal}(a), the red plot is barely visible below the green, which shows that the thickness is sufficiently large to yield the expected energy for the bulk states. In Figs.\ \ref{fig:C4_nonideal}(b) and \ref{fig:C4_nonideal}(c), we plot the energy of the lowest-energy bulk state and its IPR, respectively. Figure \ref{fig:C4_nonideal}(d) shows the energy of the surface state, calculated by exact diagonalization of the slab Hamiltonian, while \Cref{fig:C4_nonideal}(e) shows its IPR. The comparison of Figs.\ \ref{fig:C4_nonideal}(b) and \ref{fig:C4_nonideal}(d) demonstrates that the energy of the surface state is still much smaller than the gap. The flatness of the surface band can be controlled by the ratio of the parameters $c_{1,0,0}^x$ and $c_{0,1,0}^x$ to $c_{0,0,1}^z$. The fact that the IPR in \Cref{fig:C4_nonideal}(e) is much larger than the one in \Cref{fig:C4_nonideal}(c) shows that the state close to zero energy is still much more strongly localized than the bulk state and the perturbation $\mathcal{H}^{(1)}$ does not destroy the surface state.

\section{Summary and conclusions}\label{sec:summary}

Superconductors with zero-energy flat bands of Majorana states at their surfaces accompanied by full energy gaps in the bulk would be of tremendous interest from both fundamental and applied perspectives. However, crystal symmetries and TRS generically cannot guarantee such a situation. We have shown that it should nevertheless be possible to realize such systems at least approximately, i.e., with a nearly flat surface band in a full bulk gap. We have done this in two steps: First, we have derived the conditions for ideal flat surface bands. We have identified a winding number that protects such flat bands at specific surfaces of noncentrosymmetric superconductors with TRS. For every point in the sBZ, these superconductors must have a SOC vector that does not change in direction---but may change in magnitude---as a function of the momentum component normal to the surface and is parallel to the triplet-pairing vector. Consequently, the BdG Hamiltonian commutes with a component of the electronic spin and can be block-diagonalized into spin-up and a spin-down blocks. While the fully gapped BdG Hamiltonian of the entire system obeys TRS $\mathcal{T}$ with $\mathcal{T}^2=-1$, PHS $\mathcal{C}$ with $\mathcal{C}^2=+1$, and chiral symmetry, and is therefore in symmetry class DIII, the spin-up and spin-down blocks individually obey neither TRS nor PHS but retain chiral symmetry so that they are both in symmetry class AIII. Due to the chiral symmetry, we can bring the spin blocks into off-diagonal form and define a winding number as the winding of the off-diagonal entries in the complex plane when the momentum component normal to the surface traverses the bulk Brillouin zone. We have discussed a simple model for which the winding number can be written in terms of the signs of the gaps on the spin-up and spin-down bands at the respective Fermi surfaces. If the triplet-pairing amplitude is sufficiently large and the normal-state dispersion is sufficiently flat in the directions parallel to the surface, i.e., it is quasi-one-dimensional, the chemical potential can be chosen such that the winding number is nonzero in the entire sBZ.

Furthermore, we have identified the point groups that permit such a nonzero winding number. As an example, we have calculated the surface-state and lowest bulk-state energy for the BdG Hamiltonian of a slab with $C_4$ point-group symmetry and a SOC vector pointing in the $z$ direction. We have also calculated the IPRs for these states to show that there is indeed a zero-energy surface state and a finite bulk gap everywhere in the sBZ. The method of choice for the detection of (approximately) zero-energy surface bands is angle-resolved photoemission spectroscopy (ARPES). However, ARPES may not work because heating effects could prevent reaching the required low temperatures. The key experimental signature of the zero-energy surface band is then a zero-bias conductance peak in low-temperature tunneling experiments~\cite{TMYYS10, SR11, BST11, SBT12, TRB15, SB15}.

In the second step, we have examined the effect of a small component in the SOC term that breaks the spin-component symmetry. This is the realistic situation because, in contrast to the occurrence of an exact commutation with a spin component, it does not require fine-tuning of parameters. We treat the symmetry-breaking term as a small perturbation. We have employed a quasiclassical approximation for the unperturbed surface state and have derived its energy within degenerate perturbation theory. The dispersion of the surface states and their band width turn out to be of first order in the symmetry-breaking SOC terms. We have compared these results to the numerical exact diagonalization of the BdG Hamiltonian of a slab for the exemplary point group $C_4$, finding very good agreement of the quasiclassical and perturbative approximation and exact diagonalization. The results show that sufficiently weak breaking of the spin-rotation symmetry does not destroy the surface states and keeps them at low energies but makes their dispersion weakly dispersive. On the other hand, the bulk gap remains large. We therefore expect a somewhat broadened zero-bias conductance peak in low-temperature tunneling experiments for such systems~\cite{TMYYS10, BST11, SBT12}.

Our work describes the surface bands at the BdG mean-field level, i.e., the quasiparticle states forming these bands are effectively non-interacting. Since the band width is small, residual interactions between the quasiparticles easily drive the system into a strongly correlated regime. The correlation effects will be fundamentally different from, e.g., flat bands in magic-angle twisted bilayer graphene, in that the Bogoliubov quasiparticles are not charge eigenstates and their average charge is small. Hence, interactions mediated by the Cooper-pair condensate and possibly by phonons will play an important role. The proposed setup thus provides a route to unconventional, strongly correlated, two-dimensional fermionic systems.

A reasonable next step is to search for suitable candidate materials. These materials should be superconductors with a crystal structure belonging to one of the point groups $C_1$, $C_2$, $C_3$, $C_4$, $C_6$, $D_2$, $D_3$, $D_4$, $D_6$, $C_{2v}$, or $C_s$. Moreover, they should have a Fermi surface which is sufficiently flat in directions parallel to a surface such that the projection covers the entire sBZ. Moreover, the material should have a strong triplet-to-singlet ratio for the superconducting gap. Among the identified candidates, one can then try to find the ones which come closest to a unidirectional SOC vector.

\begin{acknowledgments}

We thank P. M. R. Brydon and A. P. Schnyder for useful discussions. Financial support by the Deutsche Forschungsgemeinschaft through Collaborative Research Center SFB 1143, project A04, project id 247310070, and the Würzburg-Dresden Cluster of Excellence ct.qmat, EXC 2147, project id
390858490, is gratefully acknowledged.

\end{acknowledgments}

\appendix

\section{Conditions for nonzero winding number}
\label{sec:appenix_1}

In this appendix, we derive the conditions under which the winding number defined in Eq.~\eqref{eq:windingupdown} can be nonzero in the entire sBZ. The winding number $W_{\perp,\sigma}(\mathbf{k}_\shortparallel)$ describes the winding of the image of the function $D_\sigma(k_\perp,\mathbf{k}_\shortparallel) = \epsilon_\mathbf{k} + \sigma \lambda l(\mathbf{k}) + i\, [\Delta^s_\mathbf{k} + \sigma \Delta^t_\mathbf{k} l(\mathbf{k})]$, see Eq.~\eqref{eq:D_updown}, around the origin of the complex plane. We note the close connection between the real part of $D_\sigma(k_\perp,\mathbf{k}_\shortparallel)$ and the dispersion of the positive-helicity and negative-helicity bands $\xi^\pm_\mathbf{k}$:
\begin{equation}\label{eq:appA.Dvsxi}
\text{Re}\, D_\sigma(k_\perp,\mathbf{k}_\shortparallel)
  = \epsilon_\mathbf{k} + \sigma \lambda l(\mathbf{k})
  = \begin{cases}
    \xi^\sigma_\mathbf{k} & \text{for } l(\mathbf{k}) \geq 0,\\
    \xi^{-\sigma}_\mathbf{k} & \text{for } l(\mathbf{k})<0,
\end{cases}
\end{equation}
as well as of the imaginary parts of $D_\sigma(k_\perp,\mathbf{k}_\shortparallel)$ and the gap $\Delta^\pm_\mathbf{k}$ on the two bands,
\begin{equation}
\text{Im}\, D_\sigma(k_\perp,\mathbf{k}_\shortparallel)
  = \Delta^s_\mathbf{k} + \sigma \Delta^t_\mathbf{k} l(\mathbf{k})
  = \begin{cases}
    \Delta^\sigma_\mathbf{k} & \text{for } l(\mathbf{k}) \geq 0,\\
    \Delta^{-\sigma}_\mathbf{k} & \text{for } l(\mathbf{k})<0.
\end{cases}
\end{equation}
We assume that for every point $\mathbf{k}_\shortparallel$ in the sBZ, both $\xi^+_{k_\perp,\mathbf{k}_\shortparallel}=0$ and $\xi^-_{k_\perp,\mathbf{k}_\shortparallel}=0$ have exactly two solutions each, which we denote by $k_{\perp}^{(1)}>0$, $k_{\perp}^{(2)}<0$ and $k_{\perp}^{(3)}>0$,  $k_{\perp}^{(4)}<0$, respectively. This means that the projections of the positive-helicity and the negative-helicity Fermi surfaces both cover the entire sBZ, which can for example be achieved in a model where the hopping is much stronger in the direction orthogonal to the surface than in the in-plane directions and the SOC is sufficiently weak.

We rewrite Eq.~\eqref{eq:windingupdown} as
\begin{align}\label{eq:windingupdown_new}
&W_{\perp,\sigma}(\mathbf{k}_\shortparallel)
  = \frac{1}{2\pi} \int_{k_\perp} dk_\perp\, \partial_{k_ \perp} \notag \\
&\quad{}\times\arg\big( {-}[\Delta^s_{\mathbf{k}} + \sigma \Delta^t_\mathbf{k} l(\mathbf{k})]
  + i\, [\epsilon_{\mathbf{k}} + \sigma \lambda l(\mathbf{k})] \big) .
\end{align}
The argument function is smooth except for a branch cut on the negative real axis, where it jumps by $\pm 2\pi$, depending on the direction in which the real axis is crossed. Since the argument function is real valued and momentum space is periodic the only contributions to the integral come from points $k_{\perp,\sigma}^{0,(i)}$ where $-[\Delta^s_{\mathbf{k}} + \sigma \Delta^t_\mathbf{k}  l(\mathbf{k})] + i\, [\epsilon_{\mathbf{k}} + \sigma \lambda l(\mathbf{k})]$ crosses the negative real axis:
\begin{align}\label{eq:winding_sum}
W_{\perp,\sigma}(\mathbf{k}_\shortparallel)
  &= - \sum_{i} \Theta \left[\Delta^s_{k_{\perp,\sigma}^{0,(i)},\mathbf{k}_\shortparallel}
   + \sigma \Delta^t_{k_{\perp,\sigma}^{0,(i)},\mathbf{k}_\shortparallel}
   l\left(k_{\perp,\sigma}^{0,(i)},\mathbf{k}_\shortparallel\right) \right] \notag\\
&\quad{} \times\text{sgn}\left(\left.\frac{\partial [\epsilon_{\mathbf{k}}
  + \sigma \lambda l(\mathbf{k})]}{\partial k_\perp}
  \right|_{k_\perp=k_{\perp,\sigma}^{0,(i)}} \right) ,
\end{align}
where $\Theta$ is the Heaviside step function. The sign function determines whether the $\text{arg}$ function jumps by $+2\pi$ or by $-2\pi$. It depends on the $k_\perp$-component of the Fermi velocity. The functions $\text{Re}\,D_\sigma(k_\perp,\mathbf{k}_\shortparallel) = \epsilon_\mathbf{k} + \sigma \lambda l(\mathbf{k})$ can only have an even number of zeros as the image of $D_\sigma$ is a closed curve in the complex plane and therefore cannot cross the real axis an odd number of times.

Note that there is a one-to-one correspondence between the zeros $k_{\perp}^{(i)}$ of $\xi^\pm$ and the zeros of $\text{Re}\, D_\sigma(k_\perp,\mathbf{k}_\shortparallel)$, mediated by Eq.\ \eqref{eq:appA.Dvsxi}. If all the four values $k_{\perp}^{(i)}$ are also zeros of $\text{Re}\, D_\uparrow(k_\perp,\mathbf{k}_\shortparallel)$, then $\text{Re}\, D_\downarrow(k_\perp,\mathbf{k}_\shortparallel)$ is nonzero for all $k_\perp\in [-\pi,\pi)$. This would mean that $W_{\perp,\downarrow}(\mathbf{k}_\shortparallel)=0$ and thus $W_{\perp,\uparrow}(\mathbf{k}_\shortparallel) = W_{\perp,\downarrow}(-\mathbf{k}_\shortparallel)=0$ such that there is no nonzero winding number in the full sBZ. An analogous argument holds if all $k_{\perp}^{(i)}$ are zeros of $\text{Re}\, D_\downarrow(k_\perp,\mathbf{k}_\shortparallel)$.

The two momenta $k_{\perp}^{(1)}$ and $k_{\perp}^{(2)}$, which correspond to the positive-helicity Fermi surface, are zeros of $\text{Re}\, D_\uparrow(k_\perp,\mathbf{k}_\shortparallel)$ if $l(k_{\perp}^{(i)},\mathbf{k}_\shortparallel)>0$ and zeros of $\text{Re}\, D_\downarrow(k_\perp,\mathbf{k}_\shortparallel)$ if $l(k_{\perp}^{(i)},\mathbf{k}_\shortparallel)<0$. For the momenta $k_{\perp}^{(3)}$ and $k_{\perp}^{(4)}$ on the negative-helicity Fermi surface, the opposite is true. The sign of $l(k_\perp, \mathbf{k}_\shortparallel)$ cannot change between $k_\perp=k_{\perp}^{(1)}$ and $k_\perp=k_{\perp}^{(3)}$ and between $k_\perp=k_{\perp}^{(2)}$ and $k_\perp=k_{\perp}^{(4)}$ since this would lead to additional zeros of $\xi^\pm$. This leaves two possible cases: In the first case, both $l(k_{\perp}^{(1)}, \mathbf{k}_\shortparallel)$ and $l(k_{\perp}^{(3)}, \mathbf{k}_\shortparallel)$ are positive, whereas $l(k_{\perp}^{(2)}, \mathbf{k}_\shortparallel)$ and $l(k_{\perp}^{(4)}, \mathbf{k}_\shortparallel)$ are negative, such that $k_{\perp}^{(1)}$ and $k_{\perp}^{(4)}$ are solutions of $\text{Re}\, D_\uparrow(k_\perp,\mathbf{k}_\shortparallel)=0$ and $k_{\perp}^{(2)}$ and $k_{\perp}^{(3)}$ are solutions of $\text{Re}\, D_\downarrow(k_\perp,\mathbf{k}_\shortparallel)=0$. Without loss of generality, we can assume that $ \Delta^s_\mathbf{k}\equiv \Delta^s>0$ and $\Delta^t_{\mathbf{k}}=\Delta^t>0$. This means that the momenta on the positive-helicity Fermi surface both contribute to the sum in their respective winding number in Eq.~\eqref{eq:winding_sum} because the argument of the Heaviside step function for these momenta, i.e., $\Delta^+_\mathbf{k}$, is always positive. The contribution is either $+1$ or $-1$, depending on the sign of the $k_\perp$-component of their Fermi velocity, which is positive for $k_{\perp}^{(1)}$ and negative for $k_{\perp}^{(2)}$.

The momenta on the negative-helicity Fermi surface, i.e., $k_{\perp}^{(3)}$ and $k_{\perp}^{(4)}$, contribute to the sum if the corresponding gap $\Delta^-$ is positive, i.e., $\Delta^s/\Delta^t>|l(k_{\perp}^{(3)},\mathbf{k}_\shortparallel)|$ and $\Delta^s/\Delta^t>|l( k_{\perp}^{(4)},\mathbf{k}_\shortparallel)|$. If this is the case, then the contribution of the negative $k_\perp$-component of the Fermi velocity corresponding to $k_{\perp}^{(4)}$ cancels the contribution of the term corresponding to $k_{\perp}^{(1)}$ in $W_{\perp,\uparrow}(\mathbf{k}_\shortparallel)$ and the positive $k_\perp$-component of the Fermi velocity corresponding to $k_{\perp}^{(3)}$ cancels the contribution of the term corresponding to $k_{\perp}^{(2)}$ in $W_{\perp,\downarrow}(\mathbf{k}_\shortparallel)$, such that both winding numbers are zero. If $\Delta^s/\Delta^t< \min_{i\in\lbrace 3,4\rbrace} |l(k_{\perp}^{(i)},\mathbf{k}_\shortparallel)|$, then $k_{\perp}^{(3)}$ and $k_{\perp}^{(4)}$ do not contribute to the winding numbers and $W_{\perp,\uparrow}(\mathbf{k}_\shortparallel)=-1=-W_{\perp,\downarrow}(\mathbf{k}_\shortparallel)$.

In the second case, both $l(k_{\perp}^{(1)}, \mathbf{k}_\shortparallel)$ and $l(k_{\perp}^{(3)}, \mathbf{k}_\shortparallel)$ are negative and $l(k_{\perp}^{(2)}, \mathbf{k}_\shortparallel)$ and $l(k_{\perp}^{(4)}, \mathbf{k}_\shortparallel)$ are positive. An analogous line of reasoning leads to a nonzero winding number $W_{\perp,\uparrow}(\mathbf{k}_\shortparallel)=1=-W_{\perp,\downarrow}(\mathbf{k}_\shortparallel)$ if $\Delta^s/\Delta^t< \min_{i\in\lbrace 3,4\rbrace} | l(k_{\perp}^{(i)}, \mathbf{k}_\shortparallel)|$. In conclusion, we find nonzero winding numbers $W_{\perp,\uparrow}(\mathbf{k}_\shortparallel)$ and $W_{\perp,\downarrow}(\mathbf{k}_\shortparallel)$ in the full sBZ if and only if $\Delta^s/\Delta^t < \min_{i\in\lbrace 3,4\rbrace} |l(k_{\perp}^{(i)}, \mathbf{k}_\shortparallel)|$ and $l$ changes sign between the Fermi surfaces at positive $k_\perp^{(1,3)}$ and the Fermi surfaces at negative $k_\perp^{(2,4)}$. This is equivalent to requiring
\begin{align}
\text{sgn}\Big(\Delta^+_{k_{\perp}^{(1)},\mathbf{k}_\shortparallel}\Big)
  &=- \text{sgn}\Big(\Delta^-_{k_{\perp}^{(4)},\mathbf{k}_\shortparallel}\Big) , \\
\text{sgn}\Big(\Delta^+_{k_{\perp}^{(2)},\mathbf{k}_\shortparallel}\Big)
  &=- \text{sgn}\Big(\Delta^-_{k_{\perp}^{(3)},\mathbf{k}_\shortparallel}\Big) ,
\end{align}
which are the conditions given in Eqs.\ \eqref{eq:gapcondition1} and~\eqref{eq:gapcondition2}.

\section{Other point groups} \label{sec:appenix_2}

\begin{table*}[!htbp]
\caption{\label{tab:SOC_general}Nearest-neighbor approximation for the antisymmetric SOC vector for all noncentrosymmetric point groups which permit a nonzero winding number $W_{\perp,\sigma}(\mathbf{k}_\shortparallel)$ in the full sBZ. The constants $c_{i,j,k}^x$, $c_{i,j,k}^y$, and $c_{i,j,k}^z$ are real valued. The second column gives the most general nearest-neighbor term in the SOC vector. The third and fourth columns give all possible options for the direction of a unidirectional SOC vector and the corresponding surface orientations that can lead to a nonzero winding number $W_{\perp,\sigma}(\mathbf{k}_\shortparallel)$. The fifth column lists the conditions which must be satisfied by the coefficients $c_{i,j,k}$ so that the SOC vector can be approximately unidirectional and lead to surface modes close to zero energy in the entire sBZ. The last column gives the component $\mathbf{l}_\mathbf{k}^\shortparallel$ of the SOC vector which is parallel to $\mathbf{n}$.}
\begin{ruledtabular}
\begin{tabular}{cllllll}
\multicolumn{1}{>{\centering\arraybackslash}b{8mm}}{Point group}& \multicolumn{1}{>{\centering\arraybackslash}b{64mm}}{SOC vector (nearest-neighbor approximation), general}                                                                                                                                                                                                                                                                                                                                                                                         & \multicolumn{1}{c}{$\mathbf{n}$}                              & \multicolumn{1}{c}{$k_\perp$} & \multicolumn{1}{>{\centering\arraybackslash}b{29mm}}{Condition for $\mathbf{l_k} \approx l(\mathbf{k}) \mathbf{n}$} & \multicolumn{1}{>{\centering\arraybackslash}b{30mm}}{Component of the SOC vector parallel to $\mathbf{n}$}                                                                             \\
\colrule
$C_1$& $\begin{aligned}[t]\mathbf{l}_\mathbf{k}=~&\mathbf{c}_{1,0,0}\sin k_x +\mathbf{c}_{0,1,0} \sin k_y+\mathbf{c}_{0,0,1}\sin k_z\end{aligned}$                           & $\mathbf{c}_{1,0,0}$                      & $k_x$     & $|\mathbf{c}_{0,1,0}|, |\mathbf{c}_{0,0,1}| \ll |\mathbf{c}_{1,0,0}|$ & $\mathbf{l}_\mathbf{k}^\shortparallel=\sin k_x\: \mathbf{c}_{1,0,0}$                                                                                                \\
                          &                                                                                                                                                                                                                                                                                                                                                                                                                                              & $\mathbf{c}_{0,1,0}$                      & $k_y$     & $|\mathbf{c}_{1,0,0}|, |\mathbf{c}_{0,0,1}| \ll |\mathbf{c}_{0,1,0}|$ & $\mathbf{l}_\mathbf{k}^\shortparallel=\sin k_y\: \mathbf{c}_{0,1,0}$                                                                                                \\
                          &                                                                                                                                                                                                                                                                                                                                                                                                                                              & $\mathbf{c}_{0,0,1}$                      & $k_z$     & $|\mathbf{c}_{1,0,0}|, |\mathbf{c}_{0,1,0}| \ll |\mathbf{c}_{0,0,1}|$ & $\mathbf{l}_\mathbf{k}^\shortparallel=\sin k_z\: \mathbf{c}_{0,0,1}$                                                                                                \\
$C_2$& \multirow{3}{*}{$\begin{aligned}[t]\mathbf{l}_\mathbf{k}=~&(c_{1,0,0}^x \sin k_x+ c_{0,1,0}^x \sin k_y) \hat{\mathbf{x}} \\&+(c_{1,0,0}^y \sin k_x+ c_{0,1,0}^y \sin k_y) \hat{\mathbf{y}}\\&+c_{0,0,1}\sin k_z \hat{\mathbf{z}}\end{aligned}$}                                                                                                                                                                                                                                      & $\perp \hat{\mathbf{z}}$ & $k_x$     & $\begin{aligned}[t]&|c_{0,1,0}^x|,|c_{0,1,0}^y|, |c_{0,0,1}^z|\\& \ll  |c_{1,0,0}^x|, |c_{1,0,0}^y|\end{aligned}$   & $\begin{aligned}[t]\mathbf{l}_\mathbf{k}^\shortparallel=~&\sin k_x \\&(c_{1,0,0}^x\hat{\mathbf{x}}+c_{1,0,0}^y \hat{\mathbf{y}})\end{aligned}$                                                        \\
                          &                                                                                                                                                                                                                                                                                                                                                                                                                                              &                                           & $k_y$     & $\begin{aligned}[t]&|c_{1,0,0}^x|,|c_{1,0,0}^y|, |c_{0,0,1}^z| \\&\ll |c_{0,1,0}^x|, |c_{0,1,0}^y|\end{aligned}$   & $\begin{aligned}[t]\mathbf{l}_\mathbf{k}^\shortparallel=~&\sin k_y \\&(c_{0,1,0}^x  \hat{\mathbf{x}}+c_{0,1,0}^y \hat{\mathbf{y}})\end{aligned}$                                                       \\
                          &                                                                                                                                                                                                                                                                                                                                                                                                                                              & $\hat{\mathbf{z}}$                        & $k_z$     & $\begin{aligned}[t]&|c_{1,0,0}^x|, |c_{0,1,0}^x|, |c_{1,0,0}^y|, \\&|c_{0,1,0}^y| \ll |c_{0,0,1}^z|\end{aligned}$     & $\mathbf{l}_\mathbf{k}^\shortparallel=c_{0,0,1}\sin k_z \hat{\mathbf{z}}$                                                                                         \\
$C_3$                     & $\begin{aligned}[t]\mathbf{l}_\mathbf{k}=~& [c_{1,0,0}^x \sin(k_x/2) \cos(\sqrt{3}k_y/2)+ c_{1,0,0}^y \\&\cos(k_x/2) \sin(\sqrt{3}k_y/2) +c_{1,0,0}^x \sin k_x] \hat{\mathbf{x}}\\&-[-3 c_{1,0,0}^x \cos(k_x/2)    \sin(\sqrt{3}k_y/2)+\\&c_{1,0,0}^y \sin(k_x/2) \cos(\sqrt{3}k_y/2)+ \\&c_{1,0,0}^y \sin k_x]/\sqrt{3}~ \hat{\mathbf{y}}+ \lbrace c_{1,0,0}^z[ 1/2 \sin k_x +\\&\sin(k_x/2) \cos(\sqrt{3}k_y/2)]+ c_{0,0,1}^z\sin (k_z)\rbrace \hat{\mathbf{z}}\end{aligned}$ & $\hat{\mathbf{z}}$                        & $k_z$     & $\begin{aligned}[t]&|c_{1,0,0}^x|, |c_{1,0,0}^y| \\&\ll |c_{1,0,0}^z| \\& \ll |c_{0,0,1}^z|\end{aligned}$             & $\begin{aligned}[t]\mathbf{l}_\mathbf{k}^\shortparallel=~&\lbrace c_{1,0,0}^z[ 1/2 \sin k_x +\\&\sin(k_x/2)\\& \cos(\sqrt{3}k_y/2)]+\\& c_{0,0,1}^z\sin (k_z)\rbrace \hat{\mathbf{z}}\end{aligned}$ \\
$C_4$                     & $\begin{aligned}[t]\mathbf{l}_\mathbf{k}=~&(c_{1,0,0}^x \sin k_x+c^x_{0,1,0} \sin k_y) \hat{\mathbf{x}}+\\&(c_{1,0,0}^x \sin k_y-c^x_{0,1,0} \sin k_x ) \hat{\mathbf{y}}+\\&c_{0,0,1}^z \sin k_z  \hat{\mathbf{z}}\end{aligned}$                                                                                                                                                                                                                                                             & $\hat{\mathbf{z}}$                        & $k_z$     & $|c_{1,0,0}^x|,|c_{0,1,0}^x| \ll |c_{0,0,1}^z|$                             & $\mathbf{l}_\mathbf{k}^\shortparallel=c_{0,0,1}^z \sin k_z  \hat{\mathbf{z}}$                                                                                              \\
$C_6$                     & $\begin{aligned}[t]\mathbf{l}_\mathbf{k}=&[c_{1,0,0}^x \sin(k_x/2) \cos(\sqrt{3}k_y/2)+c_{1,0,0}^y \\&\cos(k_x/2) \sin(\sqrt{3}k_y/2) + c_{1,0,0}^x \sin k_x] \hat{\mathbf{x}}-\\&[-3 c_{1,0,0}^x \cos(k_x/2)    \sin(\sqrt{3}k_y/2)+\\& c_{1,0,0}^y \sin(k_x/2) \cos(\sqrt{3}k_y/2)+\\& c_{1,0,0}^y \sin k_x]/\sqrt{3}~ \hat{\mathbf{y}}+c_{0,0,1}^z\sin k_z \hat{\mathbf{z}}\end{aligned}$                                                                                       & $\hat{\mathbf{z}}$                        & $k_z$     & $|c_{1,0,0}^x|, |c_{1,0,0}^y| \ll |c_{0,0,1}^z|$                            & $\mathbf{l}_\mathbf{k}^\shortparallel=c_{0,0,1}^z \sin k_z \hat{\mathbf{z}}$                                                                                      \\
$D_2$& \multirow{3}{*}{$\begin{aligned}[t]\mathbf{l}_\mathbf{k}=~& c_{1,0,0}^x \sin k_x \hat{\mathbf{x}} +c_{0,1,0}^y\sin  k_y  \hat{\mathbf{y}} \\&+c_{0,0,1}^z \sin k_z \hat{\mathbf{z}}\\&~\end{aligned}$}                                                                                                                                                                                                                                                                                        & $\hat{\mathbf{x}}$                        & $k_x$     & $|c_{0,1,0}^y|, |c_{0,0,1}^z| \ll |c_{1,0,0}^x|$                            & $\mathbf{l}_\mathbf{k}^\shortparallel= c_{1,0,0}^x \sin k_x \hat{\mathbf{x}} $                                                                                             \\
                          &                                                                                                                                                                                                                                                                                                                                                                                                                                              & $\hat{\mathbf{y}}$                        & $k_y$     & $|c_{1,0,0}^x|, |c_{0,0,1}^z| \ll |c_{0,1,0}^y|$                            & $\mathbf{l}_\mathbf{k}^\shortparallel= c_{0,1,0}^y\sin  k_y  \hat{\mathbf{y}} $                                                                                            \\
                          &                                                                                                                                                                                                                                                                                                                                                                                                                                              & $\hat{\mathbf{z}}$                        & $k_z$     & $|c_{1,0,0}^x|, |c_{0,1,0}^y| \ll |c_{0,0,1}^z|$                            & $\mathbf{l}_\mathbf{k}^\shortparallel= c_{0,0,1}^z \sin k_z \hat{\mathbf{z}}$                                                                                              \\
$D_3$                     & $\begin{aligned}[t]\mathbf{l}_\mathbf{k}=~&c_{1,0,0}^x\lbrace [\sin(k_x/2) \cos(\sqrt{3}k_y/2)+ \sin k_x] \hat{\mathbf{x}}\\&+\sqrt{3}\cos(k_x/2)    \sin(\sqrt{3}k_y/2)\hat{\mathbf{y}}\rbrace+\\&\lbrace c_{1,0,0}^z[ 1/2 \sin k_x +\sin(k_x/2) \cos(\sqrt{3}k_y/2)]\\&+ c_{0,0,1}^z\sin (k_z)\rbrace \hat{\mathbf{z}}\end{aligned}$                                                                                                                                         & $\hat{\mathbf{z}}$                        & $k_z$     & $\begin{aligned}[t]&|c_{1,0,0}^x| \\&\ll |c_{1,0,0}^z| \\&\ll |c_{0,0,1}^z|\end{aligned}$                          & $\begin{aligned}[t]\mathbf{l}_\mathbf{k}^\shortparallel=~&\lbrace c_{1,0,0}^z[ 1/2 \sin k_x +\\&\sin(k_x/2) \\&\cos(\sqrt{3}k_y/2)]\\&+ c_{0,0,1}^z\sin (k_z)\rbrace \hat{\mathbf{z}}\end{aligned}$ \\
$D_4$                     & $\mathbf{l}_\mathbf{k}=c_{1,0,0}^x (\sin k_x \hat{\mathbf{x}}+\sin  k_y  \hat{\mathbf{y}})+c_{0,0,1}^z \sin k_z \hat{\mathbf{z}}$                                                                                                                                                                                                                                                                                                                     & $\hat{\mathbf{z}}$                        & $k_z$     & $|c_{1,0,0}^x| \ll |c_{0,0,1}^z|$                                         & $\mathbf{l}_\mathbf{k}^\shortparallel=c_{0,0,1}^z \sin k_z \hat{\mathbf{z}}$                                                                                               \\
$D_6$                     & $\begin{aligned}[t]\mathbf{l}_\mathbf{k}=~&c_{1,0,0}^x\lbrace [\sin(k_x/2) \cos(\sqrt{3}k_y/2)+ \sin k_x] \hat{\mathbf{x}}+\\&\sqrt{3}\cos(k_x/2)    \sin(\sqrt{3}k_y/2)\hat{\mathbf{y}}\rbrace+\\&c_{0,0,1}^z\sin (k_z)\hat{\mathbf{z}}\end{aligned}$                                                                                                                                                                                                                              & $\hat{\mathbf{z}}$                        & $k_z$     & $|c_{1,0,0}^x|\ll |c_{0,0,1}^z|$                                          & $\mathbf{l}_\mathbf{k}^\shortparallel=c_{0,0,1}^z \sin k_z \hat{\mathbf{z}}$                                                                                      \\
$C_{2v}$& $\mathbf{l}_\mathbf{k}=c_{0,1,0}^x \sin k_y\hat{\mathbf{x}}+c_{1,0,0}^y \sin k_x \hat{\mathbf{y}}$                                                                                                                                                                                                                                                                                                                                   & $\hat{\mathbf{x}}$                        & $k_y$     & $|c_{1,0,0}^y| \ll |c_{0,1,0}^x|$                                         & $\mathbf{l}_\mathbf{k}^\shortparallel=c_{0,1,0}^x \sin k_y\hat{\mathbf{x}}$                                                                                                \\
                          &                                                                                                                                                                                                                                                                                                                                                                                                                                              & $\hat{\mathbf{y}}$                        & $k_x$     & $|c_{0,1,0}^x| \ll |c_{1,0,0}^y|$                                         & $\mathbf{l}_\mathbf{k}^\shortparallel=c_{1,0,0}^y \sin k_x \hat{\mathbf{y}}$                                                                                               \\
$C_s$& $\begin{aligned}[t]\mathbf{l}_\mathbf{k}=~&c_{0,0,1}^x \sin k_z\hat{\mathbf{x}}+c_{0,0,1}^y \sin k_z \hat{\mathbf{y}}+\\&(c_{1,0,0}^z \sin k_x + c_{0,1,0}^z \sin k_y )\hat{\mathbf{z}}\end{aligned}$                                                                                                                                                                                                                                                                   & $\perp \hat{\mathbf{z}}$                  & $k_z$     & $\begin{aligned}[t]&|c_{1,0,0}^z|, |c_{0,1,0}^z| \\&\ll |c_{0,0,1}^x|, |c_{0,0,1}^y|\end{aligned}$               & $\begin{aligned}[t]\mathbf{l}_\mathbf{k}^\shortparallel=~&\sin k_z \\&(c_{0,0,1}^x \hat{\mathbf{x}}+c_{0,0,1}^y \hat{\mathbf{y}})\end{aligned}$                                                                \\
                          &                                                                                                                                                                                                                                                                                                                                                                                                                                              & $\hat{\mathbf{z}}$       & $k_x$     & $\begin{aligned}[t]&|c_{0,0,1}^x|, |c_{0,0,1}^y|, |c_{0,1,0}^z| \\&\ll |c_{1,0,0}^z|\end{aligned}$               & $\mathbf{l}_\mathbf{k}^\shortparallel=c_{1,0,0}^z \sin k_x \hat{\mathbf{z}}$                                                                                               \\
                          &                                                                                                                                                                                                                                                                                                                                                                                                                                              &                                           & $k_y$     & $\begin{aligned}[t]&|c_{0,0,1}^x|, |c_{0,0,1}^y|, |c_{1,0,0}^z| \\&\ll |c_{0,1,0}^z|\end{aligned}$               & $\mathbf{l}_\mathbf{k}^\shortparallel=c_{0,1,0}^z \sin k_y \hat{\mathbf{z}}$                                                                                               \\

\end{tabular}
\end{ruledtabular}
\end{table*}

In this appendix, we consider the effects of a not exactly unidirectional SOC vector on the surface states for systems with various point groups. There are 11 noncentrosymmetric point groups which allow for a unidirectional SOC vector and a nonzero winding number $W_{\perp,\sigma}(\mathbf{k}_\shortparallel)$ in the full Brillouin zone, namely $C_1$, $C_2$, $C_3$, $C_4$, $C_6$, $D_2$, $D_3$, $D_4$, $D_6$, $C_{2v}$, and $C_s$, as discussed in Sec.~\ref{sec:ideal_system}. Table~\ref{tab:SOC_general} shows  the nearest-neighbor approximation for the SOC vector in these groups, which is restricted according to Eq.~\eqref{eq:group_relation_l}. Moreover, the table lists all possible orientations $\mathbf{n}$ for a unidirectional SOC vector and the corresponding direction of $k_\perp$ which can lead to a nonzero winding number $W_{\perp,\sigma}(\mathbf{k}_\shortparallel)$. The last two columns list the conditions which must be fullfilled by the coefficients in order to get an approximately unidirectional SOC vector and the component $\mathbf{l}_{\mathbf{k}}^\shortparallel$ of the SOC vector which is parallel to $\mathbf{n}$.

\begin{table*}[!htbp]
\caption{\label{tab:parameters}Orientation of the SOC vector and the surface as well as parameter choices and first-order energy corrections of the surface state for the numerical calculations in Figs.\ \ref{fig:energies_point_groups_exact} and \ref{fig:difference}.}
\begin{ruledtabular}
\begin{tabular}{clcll}
Group & $\mathbf{n}$ & $k_\perp$ & \multicolumn{1}{c}{Parameters }                                                                                & \multicolumn{1}{c}{$E^{(1)}(\mathbf{k}_\shortparallel)/(\alpha_{\mathbf{k}_\shortparallel} \sqrt{(\Delta^t)^2+\lambda^2})$}                                                                                                      \\
\colrule
$C_2$       & $\hat{\mathbf{z}}$          & $k_z$     & \multicolumn{1}{>{\raggedright\arraybackslash}p{46mm}}{$c_{0,0,1}^z=1 \gg c_{1,0,0}^x=0.025$, $c_{0,1,0}^x=0.005$, $c_{1,0,0}^y=0.02$, $c_{0,1,0}^y=0.02$} & $\sqrt{(c_{1,0,0}^x \sin k_x+ c_{0,1,0}^x \sin k_y)^2+(c_{1,0,0}^y \sin k_x+ c_{0,1,0}^y \sin k_y)^2}$ \\
$C_4$       & $\hat{\mathbf{z}}$          & $k_z$     & \multicolumn{1}{>{\raggedright\arraybackslash}p{46mm}}{$c_{0,0,1}^z=1 \gg c_{1,0,0}^x=0.025$, $c_{0,1,0}^x=0.025$}                                   & $\sqrt{2}\sqrt{(c_{1,0,0}^x)^2+(c^x_{0,1,0})^2}\sqrt{\sin^2 k_x+\sin^2 k_y }$                          \\
$D_2$       & $\hat{\mathbf{z}}$          & $k_z$     & \multicolumn{1}{>{\raggedright\arraybackslash}p{46mm}}{$c_{0,0,1}^z =1 \gg c_{1,0,0}^x=0.04$, $c_{0,1,0}^y=0.03$}                                   & $\sqrt{(c_{1,0,0}^x)^2 \sin^2 k_x+(c_{0,1,0}^y)^2 \sin^2 k_y }$                                        \\
$C_s$       & $\perp\hat{\mathbf{z}}$     & $k_z$     & \multicolumn{1}{>{\raggedright\arraybackslash}p{51mm}}{$c_{0,0,1}^x,c_{0,0,1}^y=1/\sqrt{2} \gg c_{1,0,0}^z=0.02$, $c_{0,1,0}^z=0.03$}               & $|c_{1,0,0}^z \sin k_x + c_{0,1,0}^z \sin k_y|$                                                        \\
$C_{2v}$    & $\hat{\mathbf{y}}$          & $k_x$     & $ c_{1,0,0}^y=1 \gg c_{0,1,0}^x=0.05$                                                       & $|c_{0,1,0}^x \sin k_y|$          \\
\end{tabular}
\end{ruledtabular}
\end{table*}

We now use the SOC vectors from Table~\ref{tab:SOC_general} to calculate the surface-state energies in a non-ideal system, i.e., a system where the SOC vector is not exactly unidirectional for the examples of the $C_2$, $C_4$, $D_2$, $C_s$, and $C_{2v}$ groups. We require the spin-rotation symmetry in Eq.~\eqref{eq:spin_symmetry} to be approximately satisfied, which leads to certain conditions shown in the fifth column of Table~\ref{tab:SOC_general}. For the numerical calculations, we choose the parameters in such a way that these conditions are satisfied; our choice is given in Table\ \ref{tab:parameters}. Here, we do not list the point group $C_1$ as it requires many parameters to be fine tuned but does not yield any additional insight. We also leave out choices that lead to SOC vectors or first-order energy corrections which have the same form as the ones we have already listed. The first-order approximation for the surface-state energy according to Eq.~\eqref{eq:E1correction} is listed in the last column of Table~\ref{tab:parameters}. For the normal-state dispersion, we choose
\begin{equation}
\epsilon_\mathbf{k} = -\mu-2 t_\perp \cos k_\perp
  -2 t_{\shortparallel} (\cos k_{\shortparallel,1} +\cos k_{\shortparallel,2}) ,
\end{equation}
where $k_{\shortparallel,1}$ and $k_{\shortparallel,2}$ are two orthogonal momentum components parallel to the surface. We choose the parameters $t_\perp=1$, $t_\shortparallel=0.1$, $\mu=-1$,  $\lambda=0.1$, $\Delta^s=0.1$, and $\Delta^t=0.2$, which would ensure a nonzero winding number in the entire sBZ for the ideal system.

\begin{figure*}[!htbp]
\centering
\includegraphics[width=\textwidth]{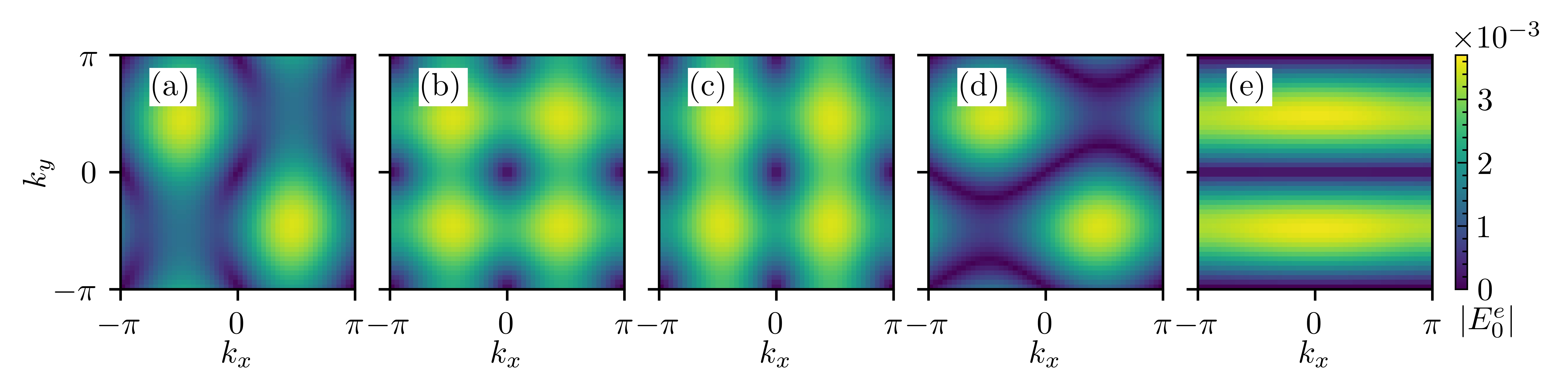}
\caption{Surface-state energy $E_0^e$ calculated by exact diagonalization of the slab BdG Hamiltonian for the point groups (a) $C_2$, (b) $C_4$, (c) $D_2$, (d) $C_s$, and (e) $C_{2v}$.}
\label{fig:energies_point_groups_exact}
\end{figure*}

The results for the surface state energies for the various point groups calculated by exact diagonalization of a slab with a thickness of $Z=500$ layers are shown in \Cref{fig:energies_point_groups_exact}. The figure shows that for all the point groups, the surface bands become weakly dispersive. The maximal surface state energy is approximately the same for all cases as it only depends on the parameters of the unperturbed system, which are chosen identically for all the point groups, and on the strength of the perturbative term, which can be measured by $\max_{\mathbf{k}_\shortparallel\in\, \text{sBZ}} |\mathbf{l}^\perp_{\mathbf{k}_\shortparallel}|$. We have chosen the parameters in the SOC vector such that this measure stays the same, namely $\max_{\mathbf{k}_\shortparallel\in\, \text{sBZ}} |\mathbf{l}^\perp_{\mathbf{k}_\shortparallel}|=0.05$. The exact shape of the surface band depends on the point group, which is also reflected by the equations for the first-order perturbative approximation given in the last column of Table~\ref{tab:parameters}.

\begin{figure*}[!htbp]
\centering
\includegraphics[width=\textwidth]{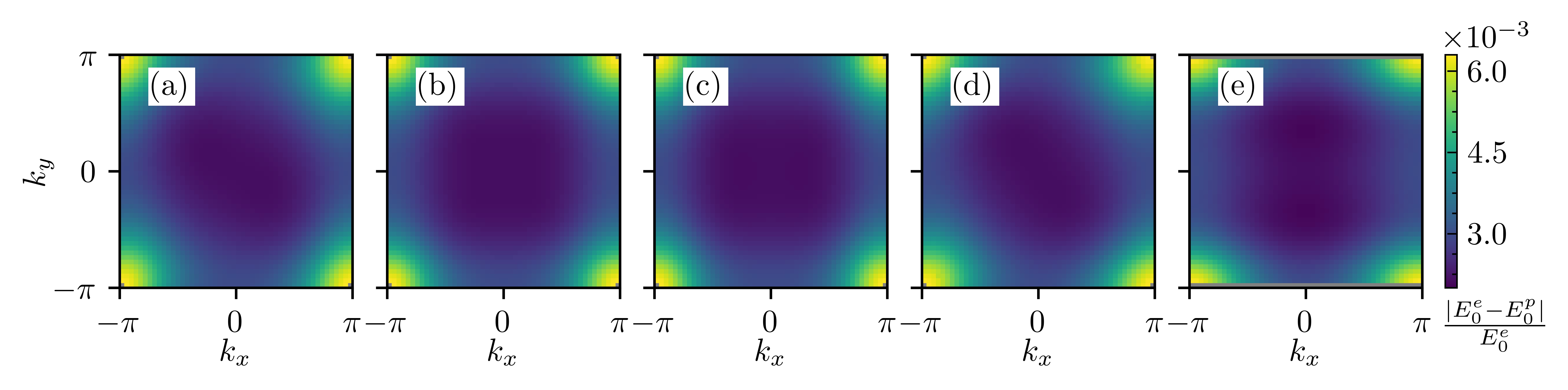}
\caption{Relative difference between the surface-state energy $E_0^e$ calculated by exact diagonalization of the slab BdG Hamiltonian and the surface-state energy $E_0^p$ calculated within perturbation theory with Eq.\ \eqref{eq:E1correction} for the point groups (a) $C_2$, (b) $C_4$, (c) $D_2$, (d) $C_s$, and (e) $C_{2v}$.}
\label{fig:difference}
\end{figure*}

Figure~\ref{fig:difference} shows the relative difference $|E_0^e-E_0^p|/E_0^e$ between the surface-state energy $E_0^e$ calculated by exact diagonalization of the slab Hamiltonian and the approximation $E_0^p$ calculated by first-order perturbation theory. All the relative differences are very small, which confirms the very good agreement between the results of exact diagonalization and the approximation in Eq.\ \eqref{eq:E1correction}.
We have also checked that there are no qualitative differences for the IPR of the surface states and the energy and IPR of the first bulk state compared to the point group $C_4$ so that they all look similar to the ones shown in \Cref{fig:C4_nonideal}.

\section{Hamiltonian matrix and exact diagonalization}
\label{sec:appendix_3}

In this appendix, we present the derivation of the Hamiltonian matrix in real space that needs to be diagonalized and comment on the numerical method. To derive the Hamiltonian of a slab of thickness $Z$, we first Fourier transform the Hamiltonian matrix in Eq.\ (\ref{eq:hamiltonian}) in the direction normal to the slab, i.e., with respect to $k_\perp$, to real space. Here, we use the example from Sec.~\ref{sec:ideal_system}, with the normal-state dispersion
\begin{equation}
\epsilon_\mathbf{k}=-\mu-2 t_z \cos k_z-2 t_{xy} (\cos k_x +\cos k_y)
\end{equation}
and the SOC vector
$\mathbf{l_k}=\sin k_\perp\: \mathbf{n}$.
For the sake of concreteness, we choose $k_\perp=k_z$ and $\mathbf{n}=\hat{\mathbf{z}}$.
The Fourier transform of the fermionic operators is defined as
\begin{equation}
c_{\mathbf{k},\sigma} = \frac{1}{\sqrt{Z}}
  \sum_{z} e^{-ik_z z}\, c_{(z,\mathbf{k}_\shortparallel),\sigma},
\end{equation}
where $c_{(z, \mathbf{k}_\shortparallel),\sigma}^\dagger$ ($c_{(z, \mathbf{k}_\shortparallel),\sigma}$) is the creation (annihilation) operator of an electron with surface momentum $\mathbf{k}_\shortparallel$ and spin $\sigma\in\lbrace\uparrow, \downarrow\rbrace$ in layer $z$.
Performing the Fourier transformation and restricting the Hamiltonian to layers $z\in\lbrace 0,\dots,Z-1\rbrace$ leads to
\begin{equation}\label{eq:Hamiltonian_full}
\mathcal{H}=\frac{1}{2}\sum_{\mathbf{k}_\shortparallel} \Psi_{\mathbf{k}_\shortparallel}^\dagger \mathcal{H}(\mathbf{k}_\shortparallel)\Psi_{\mathbf{k}_\shortparallel},
\end{equation}
with the spinors
\begin{align}
\Psi_{\mathbf{k}_\shortparallel}&=\Big(c_{(0, \mathbf{k}_\shortparallel),\uparrow},c_{(0, \mathbf{k}_\shortparallel),\downarrow}, c_{(0, -\mathbf{k}_\shortparallel),\uparrow}^\dagger,c_{(0, -\mathbf{k}_\shortparallel),\downarrow}^\dagger,\notag \\
&\qquad \dots, \notag \\
&\qquad c_{(Z-1, \mathbf{k}_\shortparallel),\uparrow},c_{(Z-1, \mathbf{k}_\shortparallel),\downarrow}, c_{(Z-1, -\mathbf{k}_\shortparallel),\uparrow}^\dagger, \notag \\
&\qquad  c_{(Z-1, -\mathbf{k}_\shortparallel),\downarrow}^\dagger\Big)^{\!T}
\end{align}
with $4Z$ components. The BdG Hamiltonian $\mathcal{H}(\mathbf{k}_\shortparallel)$ in Eq.~\eqref{eq:Hamiltonian_full} is given by the block-band matrix
\begin{equation}\label{eq:BdG_Hamiltonian_matrix}
\mathcal{H}(\mathbf{k}_\shortparallel) = \begin{pmatrix}
d^{(0)}_{\mathbf{k}_\shortparallel}&d^{(1)}_{\mathbf{k}_\shortparallel}&0 &\dots&0\\
\big(d^{(1)}_{\mathbf{k}_\shortparallel}\big)^\dagger& d^{(0)}_{\mathbf{k}_\shortparallel}& d^{(1)}_{\mathbf{k}_\shortparallel}& &\vdots\\
0&\big(d^{(1)}_{\mathbf{k}_\shortparallel}\big)^\dagger& d^{(0)}_{\mathbf{k}_\shortparallel}& \ddots & 0\\
\vdots & &\ddots &\ddots &d^{(1)}_{\mathbf{k}_\shortparallel}\\
0&\dots&0& \big(d^{(1)}_{\mathbf{k}_\shortparallel}\big)^\dagger& d^{(0)}_{\mathbf{k}_\shortparallel}
\end{pmatrix}
\end{equation}
with the $4\times4$ blocks
\vfill
\begin{widetext}
\begin{equation}
d^{(0)}_{\mathbf{k}_\shortparallel} = \begin{pmatrix}
-\mu-2 t_{xy}( \cos k_x+ \cos k_y) & 0 & 0 & \Delta^s\\
0 &-\mu -2 t_{xy}( \cos k_x+ \cos k_y)& -\Delta^s& 0\\
0 & -\Delta^s&\mu + 2 t_{xy}( \cos k_x+ \cos k_y) & 0\\
\Delta^s & 0 &0 & \mu + 2 t_{xy}( \cos k_x+ \cos k_y)
\end{pmatrix}
\end{equation}
\end{widetext}
and
\begin{equation}
d^{(1)}_{\mathbf{k}_\shortparallel} =\begin{pmatrix}
- t_z -i \lambda/2 &0 & 0 & -i \Delta^t/2\\
0 & - t_z -i \lambda/2& -i \Delta^t/2& 0\\
0 & -i \Delta^t/2& t_z -i \lambda/2 &0\\
-i \Delta^t/2& 0 &0 &  t_z +i \lambda/2
\end{pmatrix}.
\end{equation}
The matrix in Eq.~\eqref{eq:BdG_Hamiltonian_matrix} has to be diagonalized in order to obtain the eigenvalues and eigenstates of the slab. In our case, we only need the eigenvalues of lowest magnitude and the corresponding eigenstates. We have performed the diagonalization with the SciPy function {\ttfamily scipy.sparse.linalg.eigsh}, which uses the implicitly restarted Lanczos method. However, since the matrix size is only $4 Z\times 4 Z$ for slab width $Z$ (we take $Z=500$) other methods should also work.

\bibliography{Lapp}

\end{document}